\documentclass[12pt]{article}
\usepackage[utf8]{inputenc}
\usepackage[english]{babel}
\usepackage{amssymb,amsmath,graphicx}
\usepackage{multirow}
\usepackage[round,authoryear]{natbib}
\usepackage[table]{xcolor}
\usepackage{microtype}
\usepackage[hidelinks]{hyperref}
\usepackage{enumitem}
\usepackage{flafter}
\usepackage{setspace}

 \bibpunct[, ]{(}{)}{,}{a}{}{,}%
 \def\newblock{\ }%
 
\DeclareGraphicsExtensions{.pdf, .png}

\newcommand{\myblue}[1]{#1}

\title{A concise guide to existing and emerging vehicle routing problem variants}

\author{Thibaut Vidal, Peter Matl, Gilbert Laporte}

\begin{document}

\begin{center}

\vspace*{1cm}

\begin{huge}
A concise guide to existing and emerging vehicle\vspace*{0.25cm}\linebreak routing problem variants
\end{huge}

\vspace*{1.2cm}

\textbf{Thibaut Vidal$^*$} \\
Departamento de Inform\'{a}tica, Pontif\'{i}cia Universidade Cat\'{o}lica do Rio de Janeiro, Brazil \\
vidalt@inf.puc-rio.br \\
\vspace*{0.25cm}

\textbf{Gilbert Laporte} \\
CIRRELT, Canada Research Chair in Distribution Management and HEC Montréal, Canada \\
gilbert.laporte@cirrelt.ca \\
\vspace*{0.25cm}

\textbf{Piotr Matl}  \\
Department of Business Decisions and Analytics, University of Vienna, Austria \\
piotr.matl@univie.ac.at \\

\vspace*{1.5cm}


\end{center}
\noindent
\textbf{Abstract.}
Vehicle routing problems have been the focus of extensive research over the past sixty years, driven by their economic importance and their theoretical interest. The diversity of applications has motivated the study of a myriad of problem variants with different attributes. In this article, we provide a concise overview of existing and emerging problem variants. Models are typically refined along three lines: considering more relevant objectives and performance metrics, integrating vehicle routing evaluations with other tactical decisions, and capturing fine-grained yet essential aspects of modern supply chains. We organize the main problem attributes within this structured framework. We discuss recent research directions and pinpoint current shortcomings, recent successes, and emerging challenges.\\


\vfill

\noindent
$^*$ Corresponding author

\thispagestyle{empty}
\clearpage
\setcounter{page}{1}

\newpage

\section{Introduction}

Vehicle routing problems (VRPs) have been the subject of intensive and fast-growing research over sixty years. This is due to their economic importance and their theoretical interest. Using efficient vehicle routes provides a direct competitive advantage to transportation companies, which usually operate with  limited profitability margins. Moreover, the fact that these problems share a simple yet rich structure, generalizing the traveling salesman problem, has helped to elevate the VRP family into one of the main testbeds for studies in combinatorial optimization and heuristics. The VRP family can be seen as combinatorial in two senses: 1) because of the number of possible solutions, which grows exponentially with the size of the instances, and 2) because the number of conceivable problem variants also grows exponentially with the variety of problem attributes, i.e., the specific constraints, decision sets and objectives arising from real applications \citep{Vidal2012a}.

The VRP research landscape has dramatically evolved over the past two decades. Up to the early 2000s, most methodological studies were centered around a limited subset of operational problems with attributes such as time windows, multiple depots, multiple periods, and heterogeneous fleets. Since then, the number of problem variants has grown rapidly, reflecting the diversity of applications. Vehicle routing algorithms are no longer used only to produce daily routes but also serve as evaluation tools for other strategic and tactical decisions such as facility location, fleet sizing, production, and inventory management \citep{Andersson2010,Hoff2010}.

The goal of this article is to draw a succinct picture of current research in the field of vehicle routing. It is addressed to researchers and practitioners who wish to consult a concise review of existing problem features and applications. We discuss within a structured framework the main problem attributes and research directions in the field of vehicle routing as of 2020, pinpointing current shortcomings, recent successes and emerging challenges. \myblue{Given the breadth of the field, a description of every available study is now impractical. This paper therefore does not claim to be exhaustive in its coverage. Instead, we have opted for a structure based on themes rather than on VRP variants, as is the case of several books or review papers, and refer to the books of \cite{Golden2008} and \cite{Toth2014} for a more detailed coverage of specific problem variants.} This work is organized according to application-centered goals and concerns. From a high-level perspective, a VRP model can be extended along three main lines: 1) considering relevant side metrics, objectives, or combinations of objectives; 2) integrating routing optimization with other business decisions; 3) progressing toward more precise and fine-grained models.

We discuss the academic problem variants and studies according to these three classifications in Sections \ref{sec:objectives} to \ref{sec:finegrained}. Then, we highlight some important challenges and conclude in Section \ref{sec:conclusion}.

\section{Emerging Objectives -- Measuring as a Step Toward Optimizing}
\label{sec:objectives}

Measurement and quantification are central to any optimization algorithm for business processes. Most of the VRP literature considers cost as the main objective, but this does not capture all relevant performance criteria and metrics arising in practice, and many solutions based on cost optimization alone turn out to be impossible to apply in practice. In these contexts, other metrics must be considered, either as additional objectives or as constraints. We subdivide these metrics into seven main categories:

\begin{enumerate}[nosep]
\item[1)] \textbf{profitability:} \myblue{performance ratios}, profits, \myblue{outsourcing};

\item[2)] \textbf{service quality:} \myblue{cumulative} objectives, inconvenience measures, service levels;

\item[3)] \textbf{equity:} workload balance, service equity, collaborative planning;

\item[4)] \textbf{consistency:} temporal, person-oriented,  regional, \myblue{or delivery} consistency, \myblue{inconsistency};

\item[5)] \textbf{simplicity:} compactness, separation, navigation complexity;

\item[6)] \textbf{reliability:} expected cost or loss, probability of failure;

\item[7)] \textbf{externalities:} emissions, safety risks.
\end{enumerate}
This section will discuss each of these criteria and the related VRP variants. \myblue{For some applications, multiple criteria may appear as objectives (using a weighted sum, hierarchical or multi-objective formulation) or as constraints.} We analyze how each criterion has been integrated in academic problems, citing key methodological contributions and case studies.


\subsection{Profitability}

It is safe to say that \myblue{profitability or} cost optimization is the primary concern in the overwhelming majority of VRP studies. Most articles consider the minimization of total routing costs, \myblue{which may include a fixed cost per route (e.g., vehicle cost, insurance, daily wages) as well as variable costs proportional to distance or travel duration (e.g., fuel consumption, maintenance costs, hourly wages). Moreover, as outlined below, profitability also extends beyond~operational~costs.}

\paragraph{Performance ratios.}
In some situations, the optimization of routing costs is not meaningful and can even be counter-productive if it is not balanced with other performance measures. Especially for problems posed on a rolling horizon, there is a need to consider short-term surrogate objectives that approximate long-term performance goals. A practical example is the class of inventory-routing problems, for which several authors have emphasized the need to optimize the \emph{logistic ratio}. This is the ratio of routing cost to delivered quantity over the planning horizon \citep{Song2007,Benoist2011,Archetti2017b}, which measures the average cost to deliver one unit of product. This objective prevents myopic behavior that could arise from pure cost minimization \citep{Archetti2017b}. Another practical example can be found in mobility-on-demand services and in the maximization of the occupancy rate, i.e., the ratio of total passenger travel times to total vehicle travel times \mbox{\citep{Garaix2011}}. VRPs with other fractional objectives, such as profit over time, have been studied in \mbox{\cite{Baldacci2018}}.

\paragraph{Profit.}
Cost minimization often competes with profit maximization in tactical business decisions. This is especially true when the optimizer has the authority to select some of the deliveries, giving rise to the class of VRPs with profits \citep{Archetti2014}. In most of these problems, customers are associated with individual prizes, and the objective is to maximize the total profit as the difference between collected prizes and routing costs. Other problem variants maximize profit subject to distance or time constraints. These problems are connected to numerous applications in production planning and logistics \citep{Aksen2012}, manufacturing \citep{Lopez1998}, military reconnaissance \citep{Mufalli2012}, and the design of tourist itineraries \citep{Vansteenwegen2009c}, among others.

\paragraph{Outsourcing.}
To respond to growing delivery volumes while limiting the impact of high variance in shipping patterns, many freight forwarding companies regularly outsource a portion of their business to subcontractors. This practice has led to the VRP with private fleet and common carrier \citep[see, e.g.,][]{Cote2009a}, which can be viewed as a special case of VRP with profits in which each customer's prize represents its outsourcing cost. \myblue{Several variants of this problem have recently been studied.}
\cite{Krajewska2009} discuss the impact of considering heterogeneous subcontractors and distinguish three types of direct outsourcing cost: per tour, flat rate per day, and flow-based depending on distance and weight.
\cite{Ceschia2011} consider a cost function with coefficients that depend not only on distance and load but also on geographic aspects relating to the most distant customer on a tour. \myblue{\cite{Stenger2013a} solve a variant with multiple depots, while \cite{Stenger2013,Gahm2017} and \cite{Dabia2019} consider nonlinear cost functions arising from volume discounts. Finally, \cite{Goeke2019a} design a state-of-the-art branch-and-price algorithm.}

\subsection{Service Quality}

Although operational efficiency is important, providing superior service quality helps businesses to differentiate themselves from their competitors. Furthermore, profit measures are not the primary concern in some contexts, such as humanitarian relief operations, public transportation, and home healthcare logistics.

\paragraph{Cumulative objectives.}
In the cumulative VRP \citep{Ngueveu2010a,Silva2012}, the classical total cost objective is replaced with the sum of individual arrival times at the customers. Objectives of this type can be seen as more service-focused, and they are often proposed as relevant optimization criteria for relief effort operations \citep{Campbell2008, Golden2014}. The components of a cumulative objective can also be weighted in order to further bias the route structure toward a customer-centered perspective \citep{Huang2012}. In general, cumulative objectives are more appropriate when the distribution of the arrival times, travel times, transported load, etc.~is more important than their sum.

\paragraph{Inconvenience measures.}
Service quality is particularly important for passengers transportation. Common examples include the planning of school bus routes \citep{Park2010} and dial-a-ride services organized by home healthcare providers \citep{Cordeau2007a}. Since service quality is multi-dimensional, many criteria have been proposed to measure its different aspects \citep{Paquette2012}. From an optimization perspective, it is common to introduce measures of customer inconvenience, for example by minimizing the maximum ride time; the maximum time loss defined as the difference between the ride time and the corresponding shortest possible ride; or the deviation from a preset time window in case of earliness or lateness. Importantly, the target levels of these criteria may vary for different sets of customers. For example, emerging mobility-on-demand platforms aim to satisfy different service quality thresholds for different customer segments (e.g.~business, standard, budget) \citep{Beirigo2019}.

\paragraph{Service levels.}
As a general trend worldwide, logistics activities are being increasingly outsourced to third-party logistics (3PL) service providers. Due to large volumes and unforeseen events, 3PL providers can rarely service all requests. To guarantee a certain service level, most 3PL providers establish contracts that stipulate a minimum ratio of on-time deliveries. This gives rise to the VRP with service-level constraints \citep{Bulhoes2018}. In this problem, the deliveries are partitioned into groups, and a minimum percentage of the deliveries (or delivery load) must be fulfilled for each group. \cite{Orlis2019} describe an application to automated teller machine replenishment. In this study, the service levels are treated as soft constraints, and there are penalties for non-fulfillment. Service fulfillment can even be the primary optimization objective in some home healthcare applications \citep{Rasmussen2012}.

\subsection{Equity}

Efficient solutions are not necessarily equitable. Their acceptance and implementation may be contingent on a sufficiently fair distribution of resources, responsibilities, and benefits among different stakeholders. These concerns have led to a variety of equity criteria in the routing literature, reviewed in \cite{Balcik2010} and \cite{Matl2018,Matl2018a}.

\paragraph{Workload balance.}
In routing problems in the private sector, the most common equity considerations concern \textit{internal} stakeholders, i.e., the drivers or other personnel providing the service. The aim is to balance the workload allocation in order to ensure acceptance of operational plans, to maintain employee satisfaction and morale, to reduce overtime, and to reduce bottlenecks in resource utilization. Practical examples include balancing the workload of service technicians \citep{Blakeley2003}, home healthcare professionals \citep{Liu2013}, and volunteers \citep{Goodson2014}. Balancing criteria also appear in periodic settings \citep{Groer2009,Mendoza2009} and in tactical planning problems such as service territory design \citep{Butsch2014, Kalcsics2015}. The workload $W_r$ of a route or service unit is usually operationalized through its total service duration, total demand, number of customers, or some combination of these metrics. The degree of balance is then quantified by applying an inequality function to the vector of workloads, the most common functions being minimization of the maximum workload ($\min \{\max \{W_r\}$\}) and the minimization of the range ($\min \{\max \{W_r\} - \min \{W_r\}\}$). Care should be taken when modeling equity criteria, because certain combinations of workload metrics and equity functions are not appropriate for guiding mathematical optimization methods. In particular, equity functions that are not monotonic with respect to all workloads (e.g., the range or standard deviation) can lead optimization methods to unnecessarily increase the workload (e.g., longer distance or time) of \textit{every} route in an effort to artificially satisfy the ill-posed equity criterion \citep{Matl2018}.

\paragraph{Service equity.}
In contrast to private and profit-oriented logistics businesses, public and nonprofit organizations also generally have an obligation of equitable service provision to their \textit{external} stakeholders, i.e., the users or customers \citep{Balcik2010}. The most common application areas are public transportation and humanitarian logistics. There exists a close connection between the service quality measures discussed in the previous section and service equity. In fact, the min-max constraints on service quality can also be interpreted as equity constraints. However, as discussed and analyzed by \cite{Huang2012} in the context of disaster relief, there can be discrepancies between efficacy (quality of coverage) and equity, as these issues may concern different dimensions, e.g., the quantity of supplies may satisfy the full demand at all the service points while the timeliness of delivery may be very inequitable, or vice versa. Moreover, we note that unless the quality of the worst service is tightly constrained, satisfying the corresponding min-max constraint does not imply an equitable distribution of service quality. To date, these issues have received limited attention in the VRP literature.

\paragraph{Collaborative planning.}
Due to strong competitive pressures and falling profit margins in the logistics sector, carriers have an incentive to form horizontal collaborations that pool their capacities and increase their overall efficiency \citep{Cruijssen2007a,Gansterer2018}. In such coalitions, the planning of logistics operations is performed jointly through the exchange or consolidation of transportation requests and a redistribution of costs or gains, which leads to problems of fair resource allocation and profit sharing \citep{Guajardo2016, Tinoco2017}. The collaboration should be stable in the sense that each partner's individual cost is reduced by joining the partnership and the benefit of these reductions is fairly distributed. Since many of the proposed cost allocation methods relate the distributed cost or gain to the contributed resources \citep{Guajardo2016}, the routing decisions help determining the achievable savings and the fairest benefit distribution.

\subsection{Consistency}
\label{obj:consistency}

Cost-optimal routing plans may turn out to be of limited value if they vary too much over time. Customers appreciate being served by familiar faces at regular intervals; service providers are more effective and can personalize their service when they know their customers' requirements and preferences; drivers are more efficient and drive more safely when they are familiar with the peculiarities of their routes \citep{Kovacs2014}. Establishing and maintaining these aspects of familiarity requires routing and service plans to be consistent with respect to various metrics over multiple time periods.

\paragraph{Temporal consistency.}

One aspect of consistency concerns the timing of the service provision to individual customers. The aim is to provide service at roughly the same time of day and at regular intervals. Initial studies by \cite{Groer2009}, \cite{Tarantilis2012}, and \cite{Kovacs2014} handle this feature by imposing a maximum difference between the latest and earliest arrival times at any customer location. \myblue{The resulting consistent VRP (ConVRP) is often solved by metaheuristics, since the time constraints create route interdependencies which pose considerable challenges for exact solution approaches \citep{Goeke2019}.} \cite{Feillet2014} suggest an alternative approach that discretizes the day into disjoint time segments and imposes consistency by bounding the number of different time segments during which a customer is served. Some recent works propose self-imposed time windows, whereby the service provider selects for each customer a fixed time window before the demand is known, communicates this information to the customer, and subsequently generates routing plans respecting these commitments during the planning horizon \citep{Jabali2015, Spliet2015b}. Other authors have set minimum and maximum time intervals between consecutive customer visits in periodic settings \mbox{\citep{Gaudioso1992, Coelho2013a}}.

\paragraph{Person-oriented consistency.}

Another form of consistency relates to the assignment of drivers to customers. 
If the same driver regularly visits the same customers, the quality and efficiency of the customer service improves as personal relationships become established and the driver becomes more familiar with the customers \citep{Smilowitz2013}. This type of consistency is particularly important in home healthcare logistics, where the quality of the service depends on the nurses' knowledge of the preference and needs of their patients \citep{Eveborn2006}. Personal consistency is often handled at the tactical level by creating a fixed route for each driver \citep{Christofides1971, Beasley1984} or a set of template routes that are adjusted into daily operational routes \citep{Groer2009,Tarantilis2012,Kovacs2014b}. More flexible alternatives focus on directly maximizing the number of times a unique driver visits each customer \citep{Haughton2007,Smilowitz2013}, ensuring that each driver visits at least a certain fraction of their assigned customers \citep{Spliet2016} or bounding the number of different drivers serving any customer \citep{Luo2015,Braekers2016a}.

\paragraph{Regional consistency.}

In practice, and especially in urban contexts, the efficiency of a route depends on the driver's familiarity with the addresses and buildings in the area, the typical traffic conditions on important streets or junctions, possible shortcuts or detours, etc. \citep{Holland2017}. As a result, it is desirable to maintain some form of regional consistency so that drivers can become familiar with their assigned or most common service regions and hence benefit from the associated learning effects \citep{Haughton2002,Zhong2007}. This can be seen as a generalization of person-oriented consistency, and the previously mentioned fixed routes can also be a way to delineate fixed regions \citep{Wong1984a}. Similarly, regional consistency can be enforced by constructing routes that maximize the number of visited nodes within some threshold distance to fixed master routes \citep{Sungur2010}, or maximize the number of times a driver repeatedly visits the same region \citep{Smilowitz2013}.

\paragraph{Delivery consistency.}

In contexts such as vendor-managed inventories \citep{Day2009}, it may be desirable to deliver a consistent quantity of materials or provide a consistent level of service. Since cost-minimizing solutions do not typically possess these properties, \cite{Coelho2012a} propose to constrain delivery quantities within lower and upper bounds, or to follow an order-up-to policy.

\paragraph{Inconsistency.}

Finally, \emph{inconsistency} can be desired in some applications. For cash-in-transit operations, the routes should be unpredictable from day to day to reduce the risk of robberies \citep{Bozkaya2017, Constantino2017}. For the transportation of hazardous materials, safe backup routes should be available in case of adverse weather conditions or to spread the accident risk geographically \citep{Akgun2000}. In the \textit{m}-peripatetic VRP, complete dissimilarity is ensured by requiring alternative solutions to be edge-disjoint \citep{Ngueveu2010a, Ngueveu2010c}. The \textit{k}-dissimilar VRP relaxes this constraint and minimizes the average ratio between the length of edges shared by any pair of alternative solutions and the length of the routes containing those edges \citep{Talarico2015}. \cite{Marti2009} proposed a vertex-based dissimilarity measure, maximizing for each pair of paths the average distance between the vertices in one path and their closest neighbor in the other. \cite{Zajac2016} considers geographic dissimilarity explicitly by minimizing the intersection of the geographic units visited in different solutions. \cite{Michallet2014,Hoogeboom2019} and \cite{Soriano2019} enforce temporal inconsistency by using time-window penalties when the arrival times of consecutive visits at the same customer do not differ by more than a given constant.

\subsection{Simplicity}

In complex real-life systems, the acceptance and efficient realization of vehicle routing plans often depends on their simplicity and their intuitive appeal \citep{Poot2002}. As non-experts in combinatorial optimization, drivers and dispatchers may be reluctant to trust and implement solutions that appear overly complex and counter-intuitive or that require a high degree of coordination, even if these solutions are technically optimal with respect to cost. In routing contexts, visual appeal is often synonymous with compact and non-overlapping tours \citep{Hollis2012}. This corresponds to the two fundamental notions of clustering in the machine learning literature \citep{Jain2010}: group together elements in such a way that similar elements (nearby deliveries) belong to the same cluster (compactness) and different elements (distant deliveries) belong to different clusters (separation).

\paragraph{Compactness.}
A route whose customers are geographically clustered is intuitive, because its compactness serves as a visual surrogate for low cost \citep{Rossit2019}. This intuition is typically exploited by optimization algorithms within the cluster-first, route-second category. In numerical terms, compactness is usually optimized at the route level by minimizing a measure of geographic spread. For example, \cite{Poot2002} minimize the average pairwise distance between all customers in a route, or the average distance of the customers to the route's center of gravity. A different definition of ``route center'' is proposed by \cite{Tang2006}, who define the median of a tour to be the customer that minimizes the maximum distance to any other customer in the same tour. These ideas correspond to the more general \textit{k}-means and \textit{k}-median clustering approaches. Likewise, although route compactness measures concern the operational level, they are closely connected to tactical decisions related to the design of compact distribution territories \myblue{such as those commonly used in postal deliveries (see Section 3.1).} The corresponding compactness criteria are similar, e.g., minimizing the maximum distance of a customer to their territory's (route's) center \citep{Rios2009}, the maximum distance between any pair of customers in the same territory \citep{Lin2017}, or the ratio of the territory's (route's) perimeter to the total perimeter of the service area \citep{Lei2015}. Other criteria are based on geometric ratios to ideal shapes like squares and circles \citep{Kalcsics2015} or even temporal characteristics such as time-window differences \citep{Schneider2015}.

\paragraph{Separation.}
Routes that are geographically separate make coordination easier, because local changes (e.g., unexpected demand) do not impact the rest of the plan \citep{Lum2017}. Moreover, if the geographic separation is done at the tactical level, then processes such as sorting can be executed in parallel with routing, reducing delivery times and improving competitiveness \citep{Janssens2015}. Unlike compactness, separation measures are calculated at the solution level, as they concern the relationship between multiple routes. They all minimize some measure of overlap. Example metrics include the number of customers that are closer to another route's center or median than to their own \citep{Poot2002, Tang2006}, the number of edges shared by two or more routes in an arc routing context \citep{Constantino2015}, the number of customers contained in the convex hull of a route that is not their own \citep{Poot2002}, the average overlap of the routes' convex hulls \citep{Lum2017}, the number of times different routes cross paths \citep{Poot2002}, and others \citep{Corberan2017}. Although these metrics may initially appear somewhat ad hoc, they can have meaningful properties. For example, \cite{Tang2006} show that if no customer is closer to another route's median than to its own, then the convex hulls of the routes cannot overlap.

\paragraph{Navigation complexity.}
Routes should be easy to follow and execute. Distribution companies such as UPS prefer simple route structures so that drivers spend less effort on spatial route cognition and instead concentrate on driving safely \citep{Holland2017}. Users of consumer navigation systems prefer routes that are concisely described and can be easily followed, especially when traveling through unfamiliar environments \citep{Shao2014}. In practice, metrics for quantifying the navigation complexity of a route are commonly based on the number and type of turns encountered. Turn restrictions and turn penalties frequently arise in arc routing applications \citep{Assad1995,Benavent1999,Corberan2002,Vidal2017b} and can be refined by considering different types of intersections as well as the road network hierarchy \citep{Duckham2003}.

\subsection{Reliability}

Deterministic VRPs consider that all problem information is available and accurate. However, data are always subject to approximations, and unexpected events can render ``optimal'' deterministic routing plans inefficient or impracticable. As a consequence, finding reliable routing solutions that remain effective in the presence of uncertainty has become a major concern \citep{Gendreau2014,Gendreau2016}.
Under uncertainty, a natural but cost-ineffective strategy is to use a deterministic model to generate reliable solutions that contain some slack (e.g., capacity or time). 
A better option is to exploit additional knowledge of the uncertain events, in the form of representative scenarios, probability distributions, or uncertainty sets, giving rise to stochastic or robust VRP models. Beyond a mere choice of objective function, defining a stochastic VRP requires to specify when and how stochastic parameter values are observed, and when decisions are taken. Two main groups of approaches can be distinguished: 1) stochastic programming models, which typically focus on minimizing the expected cost of the routes and recourse actions made as a consequence of uncertain events; and 2) chance constraints or robust formulations, which impose constraints on the failure probabilities.

\paragraph{Expected cost or loss.}
Stochastic models based on a priori optimization \citep{Bertsimas1990} assume that the routing decisions are made in a first stage based on partial knowledge of future events (before any stochastic parameters are observed), and that prespecified recourse policies will be used in a second stage when unexpected events occur (e.g., a direct return to the depot in the case of excess demand). Most models in this family focus on optimizing the expected cost of first-stage routes and second-stage recourse actions.
There are three main sources of uncertainty: customer demands \citep{Bertsimas1992a}, service requests \citep{Jaillet1988b}, and travel times \citep{Laporte1992}. Yet, despite considerable algorithmic progress over the last four decades, the solution methods (metaheuristics and mathematical programming methods alike) are limited by the necessity to evaluate the expected cost of the recourse actions. Therefore, strong assumptions are typically made to keep the evaluations tractable: simplistic recourse policies are used, and the probability distributions associated with the random events are assumed to be independent.
Ongoing research is exploring more sophisticated recourse policies \citep{Yang2000b,Ak2007,Louveaux2018,Salavati-Khoshghalb2019}, correlated random events \citep{Rostami2017}, and multiple decision stages \citep{Dror1989,Goodson2013}.

\paragraph{Risk of failure.}
Models based on chance constraints or robust optimization impose constraints on the probability of failure as opposed to optimizing the expected cost of the uncertain events. These approaches significantly differ in how they model stochastic parameters. Chance constraints still rely on distributional information to evaluate and bound the probabilities of failure. This paradigm has been commonly used to solve VRPs with stochastic travel times and time windows \citep{Laporte1992,Li2010}. However, as highlighted in \cite{Errico2018}, there is a thin line between model assumptions that allow for efficient calculations (e.g., convolutions and dominance properties) and those that lead to intractable problems. In contrast, robust models rely on an uncertainty set (e.g., a polytope) to represent reasonable parameter variations and seek solutions that are feasible for any parameter realization within this set \citep{Ben-Tal2009,Bertsimas2011}. Robust models are especially useful in situations where no complete distribution information is available, and they are typically easier to solve than their chance-constrained counterparts \citep{Sungur2008,Gounaris2013,Pessoa2018b}. Since these models are completely risk-averse, research continues on alternative models of uncertainty (e.g., distributionally robust models) that are meaningful in practice and remain tractable \citep{Jaillet2016a,Zhang2019a}.

\subsection{Externalities}

Although transportation is essential for modern businesses and society, it also has undesirable consequences \citep{Demir2015}. A more holistic optimization of logistics and mobility is needed to mitigate the impacts of externalities while  maintaining efficient transportation systems.

\paragraph{Emissions.}
Road transportation is a major contributor to increasing atmospheric pollution caused by greenhouse gases and particulates. Reflecting also the broader societal concerns about sustainability, the past decade has seen a rapid growth in studies falling under the class of \emph{green} VRPs that account for emissions in the optimization model \citep{Demir2014a}. It has indeed been recognized that classical cost-minimizing objectives (in terms of distance or time) do not lead to minimal emissions or fuel consumption, although there is a correlation \citep{Bektas2011a}. A variety of fuel consumption models and solution methods have therefore been put forward \citep{Demir2011,Demir2014a,Kramer2015,Fukasawa2018}. Due to the complexity of the emissions functions, optimization methods need to handle various factors, e.g., load-dependency \citep{Kara2007}, time-dependency \citep{Jabali2012, Franceschetti2013}, heterogeneous fleets \citep{Koc2015}, and modal choice \citep{Bauer2010}. Although direct speed optimization is difficult to plan for road-based operations, it is an important concern and easier to achieve in maritime transportation \citep{Fagerholt2009,Norstad2011,Hvattum2013}. From a practical perspective, it is worth noting that by allowing a small increase in distance or time, one can significantly reduce emissions, which motivates the consideration of fuel consumption as a side objective \citep{Demir2014}.

\paragraph{Safety Risks.}
When transporting hazardous materials (hazmat) such as nuclear waste, chemical agents, or noxious gases, risk mitigation is a priority. Since the degree of risk and the severity of a potential accident are closely related to the selected route, classical VRP models must be carefully extended to properly incorporate various aspects of risk. For example, \cite{Tarantilis2001} consider population exposure risk on each link of the network, \cite{Ma2012} propose the inclusion of link-specific risk capacities, and \cite{Taslimi2017} examine a bilevel problem in which a regulator decides which links to close for hazmat transportation while considering the expected alternative routes then chosen by the hazmat carriers. Accident risk is also considered along the temporal dimension by \cite{Meng2005} and \cite{Toumazis2013}, who consider time-dependent risk models, and by \cite{Zografos2004}, who examines the trade-off between travel time and risk. Note that some hazmat VRP models can be generalized to different types of undesirable externalities, such as noise, disturbance, and pedestrian safety \citep{Bronfman2015, Grabenschweiger2018}. Finally, consumer-oriented routing applications optimize safety from the opposite perspective, aiming to generate routes that are safe for the user \citep{Shah2011,Kim2014}.

\section{Integrated Problems -- Routing as an Evaluation Tool}
\label{sec:integrated}

Vehicle routing decisions are fundamentally operational but are often linked with other decisions taken at a strategic or tactical level over a longer planning horizon \citep{Crainic2002}. In such contexts, generating VRP solutions or at least evaluating their characteristics becomes essential \myblue{to evaluate the cost of planning decisions made at a higher level, which can be districting, facility location, fleet composition, or inventory and production management.} Two main approaches are typically used: continuous approximation and regression models \citep{Franceschetti2017}, or fast versions of VRP algorithms adapted to stochastic or scenario-based problem variants. The former aims to give a good estimate of the routing costs based on geometric considerations, while the latter samples demand patterns resulting from distribution or scenario information. While stochastic and scenario-based approaches offer greater precision, they generally lead to large-scale integrated problems which challenge the capabilities of current solution methods. This section surveys the main applications and methods arising in \myblue{integrated two-level problems of which routing is one of the components}.
 
\subsection{Routing and Districting}
\label{sec:districting}

Districting is the process of partitioning a territory for political, administrative or commercial purposes \citep{Kalcsics2015}. The best-known application is the design of political districts \citep[see, e.g.,][]{Bozkaya2003}, but logistics applications are also common. These include the design of sales territories \citep{Skiera1998,Drexl1999,Lei2015} and distribution management applications, for example those encountered in mail delivery systems \citep{Rosenfield1992,Novaes2000,Bruno2019}. Fixed districts ensure regional consistency and facilitate delivery operations (see Section \ref{obj:consistency}). Districting plans are typically subject to hard constraints such as contiguity as well as soft constraints such as size, compactness, population balance, homogeneity, and fairness. These soft constraints, which are often nonlinear, are eventually aggregated into a multi-criteria objective function with suitable user-defined weights. The districts are often expected to change over time because of population shifts, for example. In such cases, robustness with respect to future stochastic or dynamic changes is also deemed to be a desirable property \citep{Lei2016}.

There are two main techniques for constructing districts. The most common  aggregates cells, usually called basic units, for which geographic, demographic or socio-economic data are available. It is common to define basic units as census tracts \citep{Bozkaya2003}. A second technique divides a planar area by drawing lines that define the district boundaries through geometric arguments, as in \cite{Carlsson2012} and \cite{Carlsson2013}. One obvious advantage of the first technique is that it lends itself to the use of local search-based metaheuristics in which basic units are iteratively relocated or swapped between adjacent districts and allows efficient evaluations of the objective function.

We focus on applications in which a traveling salesman problem (TSP) or VRP must eventually be solved within each district. A common case arises in the planning of sales districts, where each district is assigned to a vendor or a team of vendors. When designing the districts, one must take into account the routing cost and also ensure a level of equity between the routes of different districts. If, as is usually the case, a local search technique is used to optimize the districts, it can be prohibitively long to optimize the vehicle routes associated with the districts at each step (i.e., move evaluation). To circumvent this issue, most solution methods rely on closed-form formulas to approximate the routing costs without actually determining the routes. Two such formulas are the Beardwood-Halton-Hammersley (BHH) formula for the TSP \citep{Beardwood1959} and the \cite{Daganzo1984} formula for the VRP. The BHH formula approximates the routing cost through $n$ independently and identically distributed points in a compact area of size $A$ as $\beta \sqrt{nA}$, where $\beta$ is a constant. Appropriate constant values are provided in \cite{Applegate2011}. Combining this formula with a simple geometrical partitioning strategy, \cite{Daganzo1984} approximates the cost of a VRP solution as $2rm + 0.57 \sqrt{nA}$, where $m$ is the number of vehicle tours and $r$ is the average distance between the depot and the barycenters of the districts. The first term in this expression represents the ``line-haul'' distance to reach the districts, and the second term measures the routing costs within the districts.
Continuous approximation formulas are still being refined and generalized \citep[see, e.g.,][]{Cavdar2015,Merchan2019}, and approaches based on regression or neural networks \citep[see, e.g.,][]{Kwon1995} may soon achieve even better trade-offs between estimation accuracy and computational effort.

\subsection{Routing and Facility Location}
\label{sec:location}

Many applications require the evaluation of routing costs during the facility location decisions. This has led to the development of a vast literature dedicated to combined location and routing problems \citep{Prodhon2014,Laporte2015a,Schneider2017c}.
Facility location decisions are strategic or tactical in most applications. They concern warehouses, cross-docks, or satellite facilities in city logistics, whereas vehicle routes are operational decisions that can change dynamically over time. In these contexts, continuous approximation formulas can be used to estimate the routing cost \citep{Laporte1989,Campbell1990,Ouyang2006,Xie2015}, and facility catchment areas may be represented as polygons in a Voronoi diagram \citep{Laporte1989}. Continuous approximation methods can also be extended to integrate a variety of constraints and objectives, such as backbone costs in hub networks \mbox{\citep{Campbell2013a,Carlsson2013b,Carlsson2015}}.

Another approach for location and routing is to rely on Monte Carlo scenario generation as a basis for routing cost evaluations \citep{Klibi2010}. This approach, however, can lead to challenging scenario generation and optimization problems. This may explain why most studies on combined location and routing problems have opted for a deterministic ``single routing scenario'' approach, giving rise to the canonical location-routing problem (LRP), recently surveyed in \cite{Schneider2017c}. The LRP model is mainly relevant in contexts where the delivery routes are fixed over a long time, or where both location and routing decisions are operational, e.g., when locating transfer points between two vehicles or vehicle reception points such as temporary parking places and postal boxes \citep{Boudoin2014}.
The canonical LRP represents a challenge for exact algorithms \citep{Baldacci2011c,Contardo2014a}, since these approaches must ultimately enumerate many candidate subsets of locations. In contrast, metaheuristics currently produce good solutions for large-scale instances \citep{Schneider2019}. In the future, these methods may be extended to sophisticated settings with multiple routing scenarios in an attempt to improve the accuracy and applicability of tactical location routing models.

\subsection{Routing and Fleet Composition}
\label{sec:fleet-comp}

Tactical fleet sizing and composition problems occur across all transportation modalities, when renewing vehicles, adapting to market fluctuations, and evaluating business changes (e.g., company mergers). Fleet size adjustments can be done via long-term vehicle acquisitions and sales or short-term leasing. Typical planning horizons vary among applications: horizons are generally longer in maritime operations than in land-based transportation because of the long lifetime of ships and the large capital costs incurred \citep{Hoff2010}. As a result, maritime fleet sizing models usually consider fixed trade lanes for strategic planning \citep{Pantuso2014,Wang2018}. For land-based transportation, two main approaches are generally used to evaluate the routing costs within fleet composition models: continuous approximations or (multi-period or stochastic) heterogeneous VRP solution methods. 

Continuous approximation models stem from the observation that it is difficult to obtain accurate demand scenarios and even harder to solve the resulting VRPs. Time-consuming route evaluations can therefore be avoided by the use of approximation formulas to focus the optimization on the fleet sizing decisions \citep{Campbell1995,Jabali2012a,Franceschetti2017a,Nourinejad2017}.

Heterogeneous VRP models, in contrast, require the joint determination of vehicle types and routes. Each vehicle type may possess distinct characteristics, e.g., capacity, fixed and variable costs, customer-service restrictions, or even specific travel costs and speeds. Two canonical problems are generally distinguished: the fleet size and mix VRP (FMVRP) and the heterogeneous fixed fleet VRP (HFVRP). The FMVRP assumes that an unlimited number of vehicles of each type is available, whereas maximum limits are set in the HFVRP. As illustrated in the survey of \cite{Koc2016}, research on heterogeneous VRPs is extensive but usually focused on a single period in the presence of a fixed set of customer requests. This case corresponds to applications in which the fleet is already acquired (or rented for a short term) or where the demand is stable over a long time period. \cite{Kilby2016,Pasha2016}, and \cite{Bertoli2019} have recently extended the FMVRP to multi-period and stochastic settings, helping to bridge the gap between the heterogeneous VRP and its tactical fleet~composition~applications. 

Finally, the emergence of vehicles with alternative fuels and the growing focus on (locally) emission free deliveries have led to new fleet composition problems involving battery-powered and conventional vehicles \citep{Felipe2014,Pelletier2016,Hiermann2016}. Cities around the world are gradually restricting the vehicle types allowed in city centers. To cope with these challenges, there have been studies of fleet composition models with city center restrictions \citep{Davis2013,Franceschetti2017a,Hiermann2019a}. Another transition is taking place between transporter-managed and crowdsourced delivery systems. Crowdsourcing involves paying daily commuters and ad hoc drivers for last-mile deliveries in an effort to use their residual capacity, leading to a new generation of tactical fleet composition and multi-modal transportation problems \citep{Archetti2016,Arslan2019,Cleophas2019,Mourad2019}.

\subsection{Routing, Inventory, and Production Management}
\label{sec:inventory}

Inventory-routing problems (IRPs) arise in the context of vendor-managed inventory management in which a supplier jointly optimizes vehicle routes, delivery schedules, and quantities. The field is rooted in the work of \cite{Bell1983} and has since seen a phenomenal growth, discussed in the survey of \cite{Coelho2014}. Multiple versions of the problem exist, varying in the planning horizon (finite or infinite), the delivery structure (1-1, 1-M, M-M, or 1-M-M-1: see Section \ref{sec:cust-requests}), the routing patterns (back and forth routes or multi-customer routes), the inventory policy (maximum level or up-to-order), the inventory decisions (lost sales or backlogging), the fleet composition (homogeneous or heterogeneous), and the fleet size (single, multiple, or unconstrained). Since the planning horizons tend to be shorter in IRPs than in the other strategic problems discussed in this section, 
a larger part of the literature combines inventory management with route generation within integrated VRP models, \myblue{although continuous routing-cost approximations are also sometimes used \citep{Baller2019}.} Most models are defined on a rolling horizon, so the choice of objective is nontrivial. In particular, optimizing the logistic ratio \citep{Archetti2017b} can be better in practice than pure cost minimization. IRPs are notoriously challenging for exact methods \citep{Desaulniers2016a}, but there are efficient hybrid metaheuristics \citep[see, e.g.,][]{Archetti2017a,Chitsaz2019}.
 
As discussed in the surveys of \cite{Christiansen2013} and \cite{Papageorgiou2014}, IRPs have often been applied to maritime routing, particularly for the transportation of liquefied natural gas \citep[see, e.g.,][]{Stalhane2012,HalvorsenWeare2013,Andersson2016,Ghiami2019}.
Another important application is the transportation of perishable products \citep[see, e.g.,][]{Coelho2014a,Crama2018}.
Moreover, recent years have seen the emergence of inventory-routing problems related to the management of shared mobility systems, mostly in the case of bikes \myblue{and cars}. In these challenging problems, one must simultaneously optimize the inventory levels at the stations and the itineraries used to reposition the \myblue{shared vehicles} \mbox{\citep{Chemla2013a,Laporte2018}}.

Finally, supply chain integration extends well beyond inventory routing. As demonstrated by \cite{Chandra1993} and \cite{Chandra1994}, the joint optimization of routing, inventory management, and production can lead to substantial savings over a sequential approach. The resulting production-routing problem (PRP) aims to coordinate a production schedule with product deliveries at customer locations \citep{Adulyasak2015}. Recent algorithms and case studies are presented in \cite{Adulyasak2014,Absi2015,Neves-Moreira2019} and \cite{Qiu2019}. Ongoing research is considering integrating a wider set of supply chain decisions, e.g., assembly, production, inventory, and routing \citep{Chitsaz2019}, or production, location, and inventory \citep{Darvish2018}.

\section{Refined Problems -- Precise and Applicable Plans}
\label{sec:finegrained}

In parallel with studies that concern the integration of VRP models with other tactical supply chain decisions, significant research is being conducted to refine the models and integrate fine-grained problem attributes that can have a large impact on solution quality and feasibility. This section reviews some important problem refinements in relation to the transportation network, the drivers and vehicles, and the customer requests.

\subsection{Specificities of the Transportation Network}

\noindent
\textbf{Arc attributes.}
Transportation networks are usually characterized by multiple attributes, including driving time, driving cost (and tolls), transportation mode, attractiveness, safety, emissions, and energy consumption. In these conditions, a single best path may not be readily definable between each origin and destination, and several trade-off paths should be considered. For example, the canonical VRP with time windows has been extensively studied with the fundamental assumption that one time unit corresponds to one cost unit. In such situations without any trade-off, the search can be limited to a single shortest path for each origin and destination. However, time and cost are not directly proportional in real transportation networks: these resources can even be negatively correlated when tolls or access restrictions are imposed \citep{Reinhardt2016}. Research on this topic is fairly recent. Accounting for these effects gives rise to a class of VRPs on multi-graphs \citep{BenTicha2017,BenTicha2018,BenTicha2019,Hiermann2019a,Soriano2019} linked to critical applications in multi-modal transportation, long-haul transportation, and city logistics, among others \citep{Caramia2009,Garaix2010a}. Solution methods must jointly optimize the visit sequences and the paths between them. In the worst case, the number of trade-off paths between any two points grows exponentially. Still, empirical analyses have shown that this number remains small in practice for transportation networks with time-window constraints \citep{Muller-Hannemann2006,BenTicha2017}, and the set of paths could otherwise be heuristically restricted \citep{Hiermann2019a}.\\

\noindent
\textbf{Two-echelon structures.} \myblue{Studies on distribution networks possessing a two-echelon structure can be traced back to the work of \cite{Jacobsen1980}, in which intermediate facilities are used to transfer newspapers from large vehicles to smaller ones. Nowadays, as reviewed in \cite{Cuda2015} and \cite{Guastaroba2016}, e-retailers commonly adopt a two-echelon structure to deliver orders from distribution centers to cross-docking facilities for consolidation, and thence to customers. In city logistics, transfer points are typically located on the outskirts of cities to reduce noise, pollution and traffic \citep{Soysal2015}. Research on two-echelon VRPs is now very active since the joint optimization of two route levels and the related time constraints and synchronization issues pose substantial methodological challenges \citep{Grangier2016}. We refer to \cite{Breunig2015} and \cite{Marques2019} for state-of-the-art heuristic and exact algorithms.}\\

\noindent
\textbf{Congestion and time dependency.} Congestion is a major factor in city logistics, since it causes massive economic losses (400 billion dollars per year in the United States according to \citealt{Cookson2018}) and has numerous negative effects. As noted in the survey of \cite{Gendreau2015}, VRPs with time-dependent travel times (TDVRPs) may arise as a consequence of congestion, weather conditions, road closures, roaming targets, and other factors. TDVRPs have been the focus of extensive research, but the recent survey of \cite{Rincon2018} reports that the inadequate management of time-dependent travel times in routing software remains a major barrier to application. Time-dependent effects are commonly modeled via travel-time or travel-speed functions \citep{Gendreau2015}. Furthermore, the speed on an arc may be computed at its entry time (frozen link model) or may vary on the arc as time passes (elastic link model). In an elastic link model with strictly positive speeds, the FIFO property is always satisfied, i.e., a later departure leads to a later arrival time \citep{Orda1990}. This model has been used in the seminal work of \cite{Ichoua2003}.

Most studies on TDVRPs rely on a complete graph representation of the network in which each origin-destination pair is represented by a single link and travel time function. In practice, however, the time-dependent travel times are specific to each street or neighborhood of an urban network. To account for this, some studies have defined time-dependent speed functions at the network level \citep{Maden2009,Huang2017,Vidal2019b}. It is important to note that most existing vehicle routing heuristics can be adapted to the TDVRP under the condition that a fast mechanism is available for time-dependent travel time queries. Yet, despite the development of sophisticated quickest path algorithms \citep{Batz2013,Bast2016}, producing accurate speed predictions and performing rapid travel-time queries on large-scale networks (typically within a fraction of a millisecond) raise significant methodological challenges.\\

\noindent
\textbf{Access restrictions.} Turn restrictions, delays at intersections, tolls, and limited parking availability are a significant part of the reality of urban logistics. The inadequate management of these aspects is another important barrier to the application of routing software in practice \citep{Rincon2018}. \cite{Nielsen1998} estimate that turns and delays at intersections represent 30\% of the total transit time in cities, so an accurate model of turn restrictions is critical for mail delivery, waste collection, snow plowing, and street maintenance operations, among others \citep{Perrier2008,Irnich2008}. Likewise, an excessive number of turns in warehouse operations can lead to increased chances of vehicle tipovers, congestion, and collisions \citep{Celik2016}.

Accounting for these detailed effects is not straightforward. In the case of turn restrictions, for example, joining turn-feasible shortest paths may still lead to forbidden turns at their junctions. Solution approaches for such problem variants rely on graph transformations \citep{Clossey2001,Corberan2002,Vanhove2012} or exploit a \emph{mode selection} subproblem to optimize the arrival direction at each service location during route evaluations \citep{Vidal2017b}. Exact algorithms may require dedicated pricing and cut separation procedures to consider costs based on consecutive edge pairs \citep{Martinelli2015}. The limited amount of space in city centers also leads truck drivers to rely on double parking. Some recent studies have modeled the impact of such practices \citep{Morillo2014,Figliozzi2017}, yet parking considerations remain largely unrepresented in VRP models.

\subsection{Specificities of Drivers and Vehicles}

\noindent
\textbf{Heterogeneous vehicles and delivery modes.}
As discussed in Section \ref{sec:fleet-comp}, vehicle fleets are rarely homogeneous \citep{Pantuso2014,Koc2016}. Individual vehicle specificities (e.g., variable costs, specific equipments, or access restrictions) must often be explicitly considered to obtain accurate operational plans, giving rise to FMVRP and HFVRP variants. Many efficient metaheuristics and exact algorithms have been  proposed for these problems \citep[see, e.g.,][]{Vidal2012b,Koc2015,Pessoa2018,Penna2019}. Recent studies have extended the scope of heterogeneous VRPs to multi-modal transportation systems involving bikes, scooters, vans, as well as alternative propulsion modes  \citep{Felipe2014,Nocerino2016,Hiermann2019}. Beyond this, the recent growth of e-commerce has given rise to new distribution practices, including the use of drones. In the simplest case, drones make back-and-forth deliveries from a warehouse to customer locations. More sophisticated distribution modes involve the combined use of delivery vehicles and drones. For example, \cite{Murray2015,Dorling2017,Poikonen2017}, and \cite{Agatz2018} consider a delivery configuration in which a drone, mounted on a vehicle, detaches itself to perform deliveries while the vehicle keeps moving. The resulting problems are gradually giving rise to a rich research area.\\

\noindent
\textbf{Working hours regulations.} Hours-of-service (HOS) regulations are ubiquitous, and they should be taken into account when long-haul routes are generated for several days or weeks. Transportation companies, in particular, have the responsibility of ensuring that driving plans can be safely performed with regulatory break and rest periods. Typical HOS regulations in the United States, the European Union, Canada, and Australia impose daily and weekly rest periods as well as limits on the driving and working hours. Their numerous clauses, conditions, and exclusions make it extremely difficult to check that a compliant schedule exists, even for a fixed sequence of visits. \cite{Prescott-Gagnon2010} and \cite{Goel2012} have studied these rule sets for different countries and proposed efficient routing and scheduling algorithms. The latter study, in particular, used the optimized routing plans to compare various regulations in terms of their impact on drivers' fatigue.

HOS regulations also extend beyond classical single-driver day operations, and specific provisions exist for night work \citep{Goel2018} and team-driving \citep{Goel2019}. \cite{Schiffer2017c} recently highlighted the benefits of jointly planning rest periods and recharging actions for electric vehicles. A key challenge of HOS regulations relates to the purposeful use of optimization: transportation companies should verify that a feasible schedule exists, but most decisions on break and rest periods lie with the drivers. In such situations, a simple simulation of driver behavior may be more reliable than a full-blown optimization algorithm considering all regulatory aspects and exceptions. Beyond this, there is a thin line between regulatory aspects that can be optimized and those that should be used as a recourse when facing unforeseen events (e.g., the extended driving time defined by regulation (EC) 561/2006 should likely be kept as a recourse).\\

\noindent
\textbf{Loading constraints and compartments.} Trucks, ships, and airplanes have many specific load restrictions which must be taken into account during optimization \citep{Pollaris2015}. The papers considering these aspects are primarily classified by geometry, e.g., pallet loading \citep{Pollaris2017}, 2D packing \citep{Iori2007}, and 3D packing constraints \citep{Gendreau2006a}, but other constraints related to fragility, orientation, or equilibrium often come into play. Specialized applications such as car hauling require dedicated feasibility-checking mechanisms to ensure that a load can be feasibly placed on the truck \citep{DellAmico2015} and that axle-weight limits are respected \citep{Pollaris2017}. Many of these VRP variants share the common trait that load-feasibility checking, even for a fixed route, is an NP-hard problem. To speed up this critical evaluation step, a variety of packing heuristics, bounds, and rules may be used to directly filter some feasible or infeasible loads. Moreover, the loading constraints go well beyond the search for a feasible packing of items: some applications require precedence constraints (e.g., LIFO or FIFO) between services to make unloading possible \citep{Cordeau2009} \myblue{or integrate handling constraints for on-board load rearrangement \citep{Battarra2010a}}, while other applications, e.g., for hazardous materials or food transportation, impose incompatibility or separation constraints \citep{Battarra2009,Hamdi-Dhaoui2014}. The loading area can also be unique, split into different compartments \citep{Derigs2010}, or even separated into a truck and a trailer \citep{Villegas2013}. The trailer can be parked and retrieved to facilitate access to some customers, leading to two-echelon problem variants.\\

\noindent
\textbf{Recharging stops.}
There has been a rapid growth of research into VRP variants for battery-powered electric vehicles. Because of their limited range, early electric models often required en route recharging stops, and these intermediate stops \citep{Schiffer2019} became a defining feature of most electric VRPs (EVRP). The EVRP literature has quickly grown to take into account the numerous characteristics of real applications. Studies have been conducted on EVRPs with heterogeneous fleets and charging infrastructure \citep{Felipe2014,Hiermann2016,Hiermann2019}; more realistic energy consumption functions \citep{Goeke2015} and charging profiles \citep{Keskin2016,Montoya2017}; limited charging capacity \citep{Froger2017}; and time-dependent energy costs \citep{Pelletier2018}.
As battery technology progresses, the range of electric vehicles is becoming sufficient for daily delivery operations in metropolitan areas. Therefore, en route recharging is gradually disappearing from these applications. It may still be necessary for lightweight vehicles (e.g., drones) or vehicles performing round-the-clock operations. Also note that the limited supply of some materials (e.g., rare earths) and the lack of a good recycling process can limit the availability of large batteries \citep{Hwang2017}, so the development of a more efficient recharging infrastructure remains a plausible scenario.\\

\subsection{Specificities of Customer Requests}
\label{sec:cust-requests}

Some customer-request specificities arising in the form of customer-oriented objectives have been discussed in Section \ref{sec:objectives}. Here we discuss other aspects of customer requests that do not arise as an optimization goal but are nevertheless essential for useful routing plans.\\

\noindent
\textbf{Service types.}
VRP applications can involve very different service types, depending on the number of commodities involved and on the origin and destination points (depot or customer location). Four main types can generally be distinguished:
\begin{itemize}[nosep]
\item \textbf{1-M-1 (including 1-M and M-1).} One-to-many-to-one problems include depot-to-customer and customer-to-depot transportation as special cases. Applications of 1-M-1 services arise, e.g., in small-package delivery, where deliveries are made early in the routes and are followed by pickups later in the day \citep{Holland2017}.
\item \textbf{1-1.} One-to-one problems represent transportation settings in which each service is unique and associated with a fixed origin and destination. A typical application is taxi fleet operations \citep{Doerner2014}.
\item \textbf{M-M.} Many-to-many problems involve one or several resources at multiple locations. The goal is to move some of these resources toward the locations where they are most needed. Typical applications concern bike repositioning \citep{Chemla2013a,Bulhoes2018a} and lateral transshipments \citep{Paterson2011,Hartl2015}.
\item \textbf{1-M-M-1.} Finally, some applications may involve a combination of the M-M and 1-M-1 cases. One such problem was investigated by \cite{VanAnholt2016} in the context of the replenishment of automated teller machines: money has to be transferred from a central office to automated tellers, among these tellers, and back to the office.
\end{itemize}
As noted in \cite{Battarra2014a}, the first two categories of problems (1-M-1 and 1-1) have been extensively discussed in the VRP literature. In contrast, studies on M-M or 1-M-M-1 settings, especially with multiple commodities, are not as common. 

Applications also differ in terms of whether or not split shipments are allowed.  Split loads \citep{Archetti2012a} typically occur when transporting many units of the same commodity (e.g., bikes) or when delivering or collecting a divisible product (e.g., food or liquids). In such situations, a customer may be visited multiple times in order to fulfill its request. The resulting split delivery VRP is a relaxation of the capacitated VRP (CVRP) but is more complicated to solve. Because of a lack of an efficient route-based decomposition, due to the customer demands which act as linking constraints, exact methods \citep{Archetti2014b} struggle to optimally solve instances of a size (e.g., 50 to 100 deliveries) that can easily be handled in a canonical CVRP setting. Finally, applications involving pickups and deliveries with split loads have been considered in \cite{Nowak2008}, \cite{Sahin2013} and \cite{Haddad2018}. This setting is unexpectedly challenging: it is possible to create a family of benchmark instances for which any optimal solution requires a number of split pickups and deliveries that is an exponential function of the instance size \citep{Haddad2018}.\\

\noindent
\textbf{Time constraints.}
A wide range of VRP variants arising from the addition of time restrictions were surveyed in \cite{Vidal2015b}. Time constraints can arise as customer- or self-imposed time windows for deliveries \citep{Solomon1987,Agatz2010a,Jabali2015,Bruck2018}.
In addition, release and due dates for commodities can be imposed at the depot \citep{Cattaruzza2016,Shelbourne2017}. Both of these settings can be viewed as time-window constraints on pickup or delivery locations. Other applications impose response-time limits between a request and its fulfillment by a vehicle. This is critical for customer satisfaction in mobility-on-demand systems, or for the delivery of perishable products \citep{Pillac2013}. Finally, in dial-a-ride transportation services where passengers with distinct pickup and delivery locations share the same vehicle, ride-time constraints are typically imposed to limit detours for each customer \citep{Cordeau2007a,Paquette2012}. These constraints, however, make feasibility checks more complex. Research is ongoing into efficient solution evaluation procedures for these problems, to speed up heuristic search using preprocessing, incremental evaluations, and concatenations \citep{Tang2010,Vidal2015b,Gschwind2019}.\\

\noindent
\textbf{Skills.} 
Finally, maintenance or home care services may require specific skills. These requirements must be taken into account when assigning technicians and vehicles to tasks \citep{Cappanera2013,Paraskevopoulos2017,Xie2017}. In complex situations, several vehicles and workers with different skills and equipment \citep{Eveborn2009,Parragh2018} may be jointly requested for a single task, leading to VRPs with synchronization constraints; these are notoriously difficult to solve \citep{Drexl2012a}. Synchronization may even be imposed between different technicians at different locations, as in an application to electric network recovery studied by \cite{Goel2013}. In this setting, a local change in one route can impact the entire daily schedule, violating the synchronization constraints or delaying other~routes.

\section{Challenges and Prospects}
\label{sec:conclusion}

As we have shown, extensive research has been conducted over 60 years to better connect vehicle routing models and application cases.
This close proximity between academic research, software companies, and transportation actors has led to a multitude of successful applications \citep[see, e.g.,][]{Toth2014,Hall2018}.
Nonetheless, vehicle routing research is far from a closed topic. Technologies and business models evolve at a rapid pace. The continuing growth of e-commerce and home deliveries, increased access to on-demand transportation via mobility applications, and ongoing urbanization have put city transportation networks and supply chains under an unprecedented strain. To meet these challenges, companies and governing authorities seek true shifts of transportation paradigms rather than incremental optimizations of existing systems. These changes may be linked to new transportation modes, e.g., drones \citep{Poikonen2017} or autonomous vehicles \citep{Fagnant2015}, or to the redesign of business models and supply chains, e.g., crowdsourced deliveries \citep{Arslan2019} or the physical Internet \citep{Montreuil2011}. Regardless of the technology adopted, whereas products, drivers, and customers were typically aggregated into a route in classical VRPs, future applications will increasingly differentiate, synchronize, and optimize multiple flows associated with products, customers, and vehicles. The efficient coordination of such systems is a challenging task, and the associated rise in complexity rests on a fine equilibrium: while optimization models and their data requirements should be as sophisticated as required, they should also remain as~simple~as~possible.

With respect to methodology, the development of heuristics and mathematical programming algorithms that are simple and efficient yet general enough to cope with a wide gamut of VRPs remains a crucial topic. Significant progress has been achieved by disciplined research built on problem-structure analysis and decision-set decompositions \citep[see, e.g.,][]{Vidal2012b,Vidal2017b,Toffolo2019,Pessoa2019}. There is also a need to scale up VRP research. Current algorithms are usually evaluated on benchmark instances with a few hundred delivery points. This size could be strategically increased to thousands of visits to reflect emerging applications \citep{Uchoa2017,Arnold2019}. Multiple planning periods and scenarios should also be considered when relevant, e.g., for districting or location-routing. 

Finally, it is important to focus our energy on problem variants that are truly of methodological and practical interest. Indeed, solving a new VRP variant made up of an arbitrary combination of attributes is certainly a technical achievement, but it does not necessarily constitute a significant methodological~advance.
\myblue{Reproducibility and benchmarking are other important concerns.
Methodological issues such as over-tuning, as well as differences in coding protocols and in hardware have been raised by several researchers, but are not yet fully resolved. The fact that some flagship journals now require that codes be submitted as a condition for paper acceptance should, in all likelihood, foster the enforcement of stricter experimental standards.}

\section*{Acknowledgments}

This research was partly funded by the Canadian Natural and Engineering Research Council [grant number 2015-06189] as well as the National Council for Scientific and Technological Development [grant number 308498/2015-1], CAPES and FAPERJ [grant number E-26/203.310/2016] in Brazil. This support is gratefully acknowledged. \myblue{Thanks are also due to the referees for their valuable comments.}


\begin{thebibliography}{323}
\expandafter\ifx\csname natexlab\endcsname\relax\def\natexlab#1{#1}\fi
\expandafter\ifx\csname url\endcsname\relax
  \def\url#1{{\tt #1}}\fi
\expandafter\ifx\csname urlprefix\endcsname\relax\def\urlprefix{URL }\fi
\expandafter\ifx\csname urlstyle\endcsname\relax
  \expandafter\ifx\csname doi\endcsname\relax
  \def\doi#1{doi:\discretionary{}{}{}#1}\fi \else
  \expandafter\ifx\csname doi\endcsname\relax
  \def\doi{doi:\discretionary{}{}{}\begingroup \urlstyle{rm}\Url}\fi \fi

\bibitem[{Absi et~al.(2015)Absi, Archetti, Dauz{\`{e}}re-P{\'{e}}r{\`{e}}s, and
  Feillet}]{Absi2015}
Absi, N., C.~Archetti, S.~Dauz{\`{e}}re-P{\'{e}}r{\`{e}}s, D.~Feillet. 2015.
\newblock {A two-phase iterative heuristic approach for the production routing
  problem}.
\newblock {\it Transportation Science\/} {\bf 49}(4) 784--795.

\bibitem[{Adulyasak et~al.(2014)Adulyasak, Cordeau, and Jans}]{Adulyasak2014}
Adulyasak, Y., J.-F. Cordeau, R.~Jans. 2014.
\newblock {Formulations and branch-and-cut algorithms for multivehicle
  production and inventory routing problems}.
\newblock {\it INFORMS Journal on Computing\/} {\bf 26}(1) 103--120.

\bibitem[{Adulyasak et~al.(2015)Adulyasak, Cordeau, and Jans}]{Adulyasak2015}
Adulyasak, Y., J.-F. Cordeau, R.~Jans. 2015.
\newblock {The production routing problem: A review of formulations and
  solution algorithms}.
\newblock {\it Computers {\&} Operations Research\/} {\bf 55} 141--152.

\bibitem[{Agatz et~al.(2018)Agatz, Bouman, and Schmidt}]{Agatz2018}
Agatz, N., P.~Bouman, M.~Schmidt. 2018.
\newblock {Optimization approaches for the traveling salesman problem with
  drone}.
\newblock {\it Transportation Science\/} {\bf 52}(4) 965--981.

\bibitem[{Agatz et~al.(2011)Agatz, Campbell, Fleischmann, and
  Savelsbergh}]{Agatz2010a}
Agatz, N., A.M. Campbell, M.~Fleischmann, M.W.P. Savelsbergh. 2011.
\newblock {Time slot management in attended home delivery}.
\newblock {\it Transportation Science\/} {\bf 45}(3) 435--449.

\bibitem[{Ak and Erera(2007)}]{Ak2007}
Ak, A., A.L. Erera. 2007.
\newblock {A paired-vehicle recourse strategy for the vehicle-routing problem
  with stochastic demands}.
\newblock {\it Transportation Science\/} {\bf 41}(2) 222--237.

\bibitem[{Akg{\"{u}}n et~al.(2000)Akg{\"{u}}n, Erkut, and Batta}]{Akgun2000}
Akg{\"{u}}n, V., E.~Erkut, R.~Batta. 2000.
\newblock {On finding dissimilar paths}.
\newblock {\it European Journal of Operational Research\/} {\bf 121}(2)
  232--246.

\bibitem[{Aksen et~al.(2012)Aksen, Kaya, Salman, and Ak{\c{c}}a}]{Aksen2012}
Aksen, D., O.~Kaya, F.~S. Salman, Y.~Ak{\c{c}}a. 2012.
\newblock {Selective and periodic inventory routing problem for waste vegetable
  oil collection}.
\newblock {\it Optimization Letters\/} {\bf 6}(6) 1063--1080.

\bibitem[{Andersson et~al.(2016)Andersson, Christiansen, and
  Desaulniers}]{Andersson2016}
Andersson, H., M.~Christiansen, G.~Desaulniers. 2016.
\newblock {A new decomposition algorithm for a liquefied natural gas inventory
  routing problem}.
\newblock {\it International Journal of Production Research\/} {\bf 54}(2)
  564--578.

\bibitem[{Andersson et~al.(2010)Andersson, Hoff, Christiansen, Hasle, and
  L{\o}kketangen}]{Andersson2010}
Andersson, H., A.~Hoff, M.~Christiansen, G.~Hasle, A.~L{\o}kketangen. 2010.
\newblock {Industrial aspects and literature survey: Combined inventory
  management and routing}.
\newblock {\it Computers {\&} Operations Research\/} {\bf 37}(9) 1515--1536.

\bibitem[{Applegate et~al.(2011)Applegate, Bixby, Chv{\'{a}}tal, and
  Cook}]{Applegate2011}
Applegate, D.L., R.E. Bixby, V.~Chv{\'{a}}tal, W.J. Cook. 2011.
\newblock {\it {The Traveling Salesman Problem: A Computational Study}\/}.
\newblock Princeton University Press, Princeton.

\bibitem[{Archetti et~al.(2014{\natexlab{a}})Archetti, Bianchessi, and
  Speranza}]{Archetti2014b}
Archetti, C., N.~Bianchessi, M.G. Speranza. 2014{\natexlab{a}}.
\newblock {Branch-and-cut algorithms for the split delivery vehicle routing
  problem}.
\newblock {\it European Journal of Operational Research\/} {\bf 238}(3)
  685--698.

\bibitem[{Archetti et~al.(2017{\natexlab{a}})Archetti, Boland, and
  Speranza}]{Archetti2017a}
Archetti, C., N.~Boland, M.G. Speranza. 2017{\natexlab{a}}.
\newblock {A matheuristic for the multivehicle inventory routing problem}.
\newblock {\it INFORMS Journal on Computing\/} {\bf 29}(3) 377--387.

\bibitem[{Archetti et~al.(2017{\natexlab{b}})Archetti, Desaulniers, and
  Speranza}]{Archetti2017b}
Archetti, C., G.~Desaulniers, M.G. Speranza. 2017{\natexlab{b}}.
\newblock {Minimizing the logistic ratio in the inventory routing problem}.
\newblock {\it EURO Journal on Transportation and Logistics\/} {\bf 6}(4)
  289--306.

\bibitem[{Archetti et~al.(2016)Archetti, Savelsbergh, and
  Speranza}]{Archetti2016}
Archetti, C., M.W.P. Savelsbergh, M.G. Speranza. 2016.
\newblock {The vehicle routing problem with occasional drivers}.
\newblock {\it European Journal of Operational Research\/} {\bf 254}(2)
  472--480.

\bibitem[{Archetti and Speranza(2012)}]{Archetti2012a}
Archetti, C., M.G. Speranza. 2012.
\newblock {Vehicle routing problems with split deliveries}.
\newblock {\it International Transactions in Operational Research\/} {\bf
  19}(1-2) 3--22.

\bibitem[{Archetti et~al.(2014{\natexlab{b}})Archetti, Speranza, and
  Vigo}]{Archetti2014}
Archetti, C., M.G. Speranza, D.~Vigo. 2014{\natexlab{b}}.
\newblock {Vehicle routing problems with profits}.
\newblock P.~Toth, D.~Vigo, eds., {\it Vehicle Routing: Problems, Methods, and
  Applications\/}, chap.~10. Society for Industrial and Applied Mathematics,
  Philadelphia, 273--297.

\bibitem[{Arnold et~al.(2019)Arnold, Gendreau, and S{\"{o}}rensen}]{Arnold2019}
Arnold, F., M.~Gendreau, K.~S{\"{o}}rensen. 2019.
\newblock {Efficiently solving very large scale routing problems}.
\newblock {\it Computers {\&} Operations Research\/} {\bf 107}(1) 32--42.

\bibitem[{Arslan et~al.(2019)Arslan, Agatz, Kroon, and Zuidwijk}]{Arslan2019}
Arslan, A.M., N.~Agatz, L.~Kroon, R.~Zuidwijk. 2019.
\newblock {Crowdsourced delivery -- A dynamic pickup and delivery problem with
  ad hoc drivers}.
\newblock {\it Transportation Science\/} {\bf 53}(1) 222--235.

\bibitem[{Assad and Golden(1995)}]{Assad1995}
Assad, A.A., B.L. Golden. 1995.
\newblock {Arc routing methods and applications}.
\newblock M.~Ball, T.L. Magnanti, C.L. Monma, G.L. Nemhauser, eds., {\it
  Network Routing\/}, vol.~8. Elsevier, Amsterdam, 375--483.

\bibitem[{Balcik et~al.(2011)Balcik, Iravani, and Smilowitz}]{Balcik2010}
Balcik, B., S.M.R. Iravani, K.~Smilowitz. 2011.
\newblock {A review of equity in nonprofit and public sector: A vehicle routing
  perspective}.
\newblock J.J. Cochran, L.A. Cox, P.~Keskinocak, J.P. Kharoufeh, J.C. Smith,
  eds., {\it Wiley Encyclopedia of Operations Research and Management
  Science\/}. Wiley.

\bibitem[{Baldacci et~al.(2018)Baldacci, Lim, Traversi, and {Wolfler
  Calvo}}]{Baldacci2018}
Baldacci, R., A.~Lim, E.~Traversi, R.~{Wolfler Calvo}. 2018.
\newblock {Optimal solution of vehicle routing problems with fractional
  objective function}.
\newblock Tech. rep., ArXiv 1804.03316.

\bibitem[{Baldacci et~al.(2011)Baldacci, Mingozzi, and {Wolfler
  Calvo}}]{Baldacci2011c}
Baldacci, R., A.~Mingozzi, R.~{Wolfler Calvo}. 2011.
\newblock {An exact method for the capacitated location-routing problem}.
\newblock {\it Operations Research\/} {\bf 59}(5) 1284--1296.

\bibitem[{Baller et~al.(2019)Baller, Dabia, Dullaert, and Vigo}]{Baller2019}
Baller, A.C., S.~Dabia, W.E.H. Dullaert, D.~Vigo. 2019.
\newblock {The dynamic-demand joint replenishment problem with approximated
  transportation costs}.
\newblock {\it European Journal of Operational Research\/} {\bf 276}(3)
  1013--1033.

\bibitem[{Bast et~al.(2016)Bast, Delling, Goldberg, M{\"{u}}ller-Hannemann,
  Pajor, Sanders, Wagner, and Werneck}]{Bast2016}
Bast, H., D.~Delling, A.~Goldberg, M.~M{\"{u}}ller-Hannemann, T.~Pajor,
  P.~Sanders, D.~Wagner, R.F. Werneck. 2016.
\newblock {Route planning in transportation networks}.
\newblock L.~Kliemann, P.~Sanders, eds., {\it Algorithm Engineering: Selected
  Results and Surveys\/}. Springer, Berlin Heidelberg, 19--80.

\bibitem[{Battarra et~al.(2014)Battarra, Cordeau, and Iori}]{Battarra2014a}
Battarra, M., J.-F. Cordeau, M.~Iori. 2014.
\newblock {Pickup-and-delivery problems for goods transportation}.
\newblock P.~Toth, D.~Vigo, eds., {\it Vehicle Routing: Problems, Methods, and
  Applications\/}. SIAM, Philadelphia, 161--191.

\bibitem[{Battarra et~al.(2010)Battarra, Erdoǧan, Laporte, and
  Vigo}]{Battarra2010a}
Battarra, M., G.~Erdoǧan, G.~Laporte, D.~Vigo. 2010.
\newblock {The traveling salesman problem with pickups, deliveries, and
  handling costs}.
\newblock {\it Transportation Science\/} {\bf 44}(3) 383--399.

\bibitem[{Battarra et~al.(2009)Battarra, Monaci, and Vigo}]{Battarra2009}
Battarra, M., M.~Monaci, D.~Vigo. 2009.
\newblock {An adaptive guidance approach for the heuristic solution of a
  minimum multiple trip vehicle routing problem}.
\newblock {\it Computers {\&} Operations Research\/} {\bf 36}(11) 3041--3050.

\bibitem[{Batz et~al.(2013)Batz, Geisberger, Sanders, and Vetter}]{Batz2013}
Batz, G.V., R.~Geisberger, P.~Sanders, C.~Vetter. 2013.
\newblock {Minimum time-dependent travel times with contraction hierarchies}.
\newblock {\it Journal of Experimental Algorithmics\/} {\bf 18}(1) 1--43.

\bibitem[{Bauer et~al.(2010)Bauer, Bektaş, and Crainic}]{Bauer2010}
Bauer, J., T.~Bektaş, T.G. Crainic. 2010.
\newblock {Minimizing greenhouse gas emissions in intermodal freight transport:
  An application to rail service design}.
\newblock {\it Journal of the Operational Research Society\/} {\bf 61}(3)
  530--542.

\bibitem[{Beardwood et~al.(1959)Beardwood, Halton, and
  Hammersley}]{Beardwood1959}
Beardwood, J., J.H. Halton, J.M. Hammersley. 1959.
\newblock {The shortest path through many points}.
\newblock {\it Mathematical Proceedings of the Cambridge Philosophical
  Society\/} {\bf 55}(9) 299--327.

\bibitem[{Beasley(1984)}]{Beasley1984}
Beasley, J.E. 1984.
\newblock {Fixed Routes}.
\newblock {\it Journal of the Operational Research Society\/} {\bf 35}(1)
  49--55.

\bibitem[{Beirigo et~al.(2019)Beirigo, Schulte, and Negenborn}]{Beirigo2019}
Beirigo, B., F.~Schulte, R.R. Negenborn. 2019.
\newblock {A business class for autonomous mobility-on-demand: Modeling service
  quality in dynamic ridesharing systems}.
\newblock {\it IEEE Intelligent Transportation Systems Magazine\/} .

\bibitem[{Bektaş and Laporte(2011)}]{Bektas2011a}
Bektaş, T., G.~Laporte. 2011.
\newblock {The pollution-routing problem}.
\newblock {\it Transportation Research Part B: Methodological\/} {\bf 45}(8)
  1232--1250.

\bibitem[{Bell et~al.(1983)Bell, Dalberto, Fisher, Greenfield, Jaikumar, Kedia,
  Mack, and Prutzman}]{Bell1983}
Bell, W.J., L.M. Dalberto, M.L. Fisher, A.J. Greenfield, R.~Jaikumar, P.~Kedia,
  R.G. Mack, P.J. Prutzman. 1983.
\newblock {Improving the distribution of industrial gases with an on-line
  computerized routing and scheduling optimizer}.
\newblock {\it Interfaces\/} {\bf 13}(6) 4--23.

\bibitem[{Ben-Tal et~al.(2009)Ben-Tal, {El Ghaoui}, and
  Nemirovski}]{Ben-Tal2009}
Ben-Tal, A., L.~{El Ghaoui}, A.~Nemirovski. 2009.
\newblock {\it {Robust Optimization}\/}.
\newblock Princeton University Press, Princeton, NJ.

\bibitem[{{Ben Ticha} et~al.(2017){Ben Ticha}, Absi, Feillet, and
  Quilliot}]{BenTicha2017}
{Ben Ticha}, H., N.~Absi, D.~Feillet, A.~Quilliot. 2017.
\newblock {Empirical analysis for the VRPTW with a multigraph representation
  for the road network}.
\newblock {\it Computers {\&} Operations Research\/} {\bf 88}(1) 103--116.

\bibitem[{{Ben Ticha} et~al.(2018){Ben Ticha}, Absi, Feillet, and
  Quilliot}]{BenTicha2018}
{Ben Ticha}, H., N.~Absi, D.~Feillet, A.~Quilliot. 2018.
\newblock {Vehicle routing problems with road-network information: State of the
  art}.
\newblock {\it Networks\/} {\bf 72}(3) 393--406.

\bibitem[{{Ben Ticha} et~al.(2019){Ben Ticha}, Absi, Feillet, and
  Quilliot}]{BenTicha2019}
{Ben Ticha}, H., N.~Absi, D.~Feillet, A.~Quilliot. 2019.
\newblock {Multigraph modeling and adaptive large neighborhood search for the
  vehicle routing problem with time windows}.
\newblock {\it Computers {\&} Operations Research\/} {\bf 104}(1) 113--126.

\bibitem[{Benavent and Soler(1999)}]{Benavent1999}
Benavent, E., D.~Soler. 1999.
\newblock {The directed rural postman problem with turn penalties}.
\newblock {\it Transportation Science\/} {\bf 33}(4) 408--418.

\bibitem[{Benoist et~al.(2011)Benoist, Gardi, Jeanjean, and
  Estellon}]{Benoist2011}
Benoist, T., F.~Gardi, A.~Jeanjean, B.~Estellon. 2011.
\newblock {Randomized local search for real-life inventory routing}.
\newblock {\it Transportation Science\/} {\bf 45}(3) 381--398.

\bibitem[{Bertoli et~al.(2019)Bertoli, Kilby, and Urli}]{Bertoli2019}
Bertoli, F., P.~Kilby, T.~Urli. 2019.
\newblock {A column-generation-based approach to fleet design problems mixing
  owned and hired vehicles}.
\newblock {\it International Transactions in Operational Research, Articles in
  Advance\/} .

\bibitem[{Bertsimas et~al.(2011)Bertsimas, Brown, and
  Caramanis}]{Bertsimas2011}
Bertsimas, D., D.B. Brown, C.~Caramanis. 2011.
\newblock {Theory and applications of robust optimization}.
\newblock {\it SIAM Review\/} {\bf 53}(3) 464--501.

\bibitem[{Bertsimas(1992)}]{Bertsimas1992a}
Bertsimas, D.J. 1992.
\newblock {A vehicle routing problem with stochastic demand}.
\newblock {\it Operations Research\/} {\bf 40}(3) 574--585.

\bibitem[{Bertsimas et~al.(1990)Bertsimas, Jaillet, and Odoni}]{Bertsimas1990}
Bertsimas, D.J., P.~Jaillet, A.~Odoni. 1990.
\newblock {A priori optimization}.
\newblock {\it Operations Research\/} {\bf 38}(6) 1019--1033.

\bibitem[{Blakeley et~al.(2003)Blakeley, Bozkaya, Cao, Hall, and
  Knolmajer}]{Blakeley2003}
Blakeley, F., B.~Bozkaya, B.~Cao, W.~Hall, J.~Knolmajer. 2003.
\newblock {Optimizing periodic maintenance operations for Schindler Elevator
  Corporation}.
\newblock {\it Interfaces\/} {\bf 33}(1) 67--79.

\bibitem[{Boudoin et~al.(2014)Boudoin, Morel, and Gardat}]{Boudoin2014}
Boudoin, D., C.~Morel, M.~Gardat. 2014.
\newblock {Supply chains and urban logistics}.
\newblock J.~Gonzalez-Feliu, F.~Semet, J.-L. Routhier, eds., {\it Sustainable
  Urban Logistics: Concepts, Methods and Information Systems\/}. Springer,
  Berlin Heidelberg, 1--20.

\bibitem[{Bozkaya et~al.(2003)Bozkaya, Erkut, and Laporte}]{Bozkaya2003}
Bozkaya, B., E.~Erkut, G.~Laporte. 2003.
\newblock {A tabu search heuristic and adaptive memory procedure for political
  districting}.
\newblock {\it European Journal of Operational Research\/} {\bf 144}(1) 12--26.

\bibitem[{Bozkaya et~al.(2017)Bozkaya, Salman, and Telciler}]{Bozkaya2017}
Bozkaya, B., F.S. Salman, K.~Telciler. 2017.
\newblock {An adaptive and diversified vehicle routing approach to reducing the
  security risk of cash-in-transit operations}.
\newblock {\it Networks\/} {\bf 69}(3) 256--269.

\bibitem[{Braekers and Kovacs(2016)}]{Braekers2016a}
Braekers, K., A.A. Kovacs. 2016.
\newblock {A multi-period dial-a-ride problem with driver consistency}.
\newblock {\it Transportation Research Part B: Methodological\/} {\bf 94}
  355--377.

\bibitem[{Breunig et~al.(2016)Breunig, Schmid, Hartl, and Vidal}]{Breunig2015}
Breunig, U., V.~Schmid, R.H. Hartl, T.~Vidal. 2016.
\newblock {A large neighbourhood based heuristic for the two-echelon vehicle
  routing problem}.
\newblock {\it Computers {\&} Operations Research\/} {\bf 76} 208--225.

\bibitem[{Bronfman et~al.(2015)Bronfman, Marianov, Paredes-Belmar, and
  L{\"{u}}er-Villagra}]{Bronfman2015}
Bronfman, A., V.~Marianov, G.~Paredes-Belmar, A.~L{\"{u}}er-Villagra. 2015.
\newblock {The maximin HAZMAT routing problem}.
\newblock {\it European Journal of Operational Research\/} {\bf 241}(1) 15--27.

\bibitem[{Bruck et~al.(2018)Bruck, Cordeau, and Iori}]{Bruck2018}
Bruck, B.P., J.-F. Cordeau, M.~Iori. 2018.
\newblock {A practical time slot management and routing problem for attended
  home services}.
\newblock {\it Omega\/} {\bf 81} 208--219.

\bibitem[{Bruno et~al.(2019)Bruno, Cavola, Diglio, Laporte, and
  Piccolo}]{Bruno2019}
Bruno, G., M.~Cavola, A.~Diglio, G.~Laporte, C.~Piccolo. 2019.
\newblock {Reorganizing postal, collection operations in urban areas as a
  result of declining mail volumes – A case study in Bologna}.
\newblock Tech. rep., Kirkland.

\bibitem[{Bulh{\~{o}}es et~al.(2018{\natexlab{a}})Bulh{\~{o}}es, H{\`{a}},
  Martinelli, and Vidal}]{Bulhoes2018}
Bulh{\~{o}}es, T., M.H. H{\`{a}}, R.~Martinelli, T.~Vidal. 2018{\natexlab{a}}.
\newblock {The vehicle routing problem with service level constraints}.
\newblock {\it European Journal of Operational Research\/} {\bf 265}(2)
  544--558.

\bibitem[{Bulh{\~{o}}es et~al.(2018{\natexlab{b}})Bulh{\~{o}}es, Subramanian,
  Erdogan, and Laporte}]{Bulhoes2018a}
Bulh{\~{o}}es, T., A.~Subramanian, G.~Erdogan, G.~Laporte. 2018{\natexlab{b}}.
\newblock {The static bike relocation problem with multiple vehicles and
  visits}.
\newblock {\it European Journal of Operational Research\/} {\bf 264}(2)
  508--523.

\bibitem[{Butsch et~al.(2014)Butsch, Kalcsics, and Laporte}]{Butsch2014}
Butsch, A., J.~Kalcsics, G.~Laporte. 2014.
\newblock {Districting for arc routing}.
\newblock {\it INFORMS Journal on Computing\/} {\bf 26}(4) 809--824.

\bibitem[{Campbell et~al.(2008)Campbell, Vandenbussche, and
  Hermann}]{Campbell2008}
Campbell, A.M., D.~Vandenbussche, W.~Hermann. 2008.
\newblock {Routing for relief efforts}.
\newblock {\it Transportation Science\/} {\bf 42}(2) 127--145.

\bibitem[{Campbell(1990)}]{Campbell1990}
Campbell, J.F. 1990.
\newblock {Locating transportation terminals to serve an expanding demand}.
\newblock {\it Transportation Research Part B: Methodological\/} {\bf 24}(3)
  173--192.

\bibitem[{Campbell(1995)}]{Campbell1995}
Campbell, J.F. 1995.
\newblock {Using small trucks to circumvent large truck restrictions: Impacts
  on truck emissions and performance measures}.
\newblock {\it Transportation Research Part A: Policy and Practice\/} {\bf
  29}(6) 445--458.

\bibitem[{Campbell(2013)}]{Campbell2013a}
Campbell, J.F. 2013.
\newblock {A continuous approximation model for time definite many-to-many
  transportation}.
\newblock {\it Transportation Research Part B: Methodological\/} {\bf 54}
  100--112.

\bibitem[{Cappanera et~al.(2013)Cappanera, Gouveia, and
  Scutell{\`{a}}}]{Cappanera2013}
Cappanera, P., L.~Gouveia, M.G. Scutell{\`{a}}. 2013.
\newblock {Models and valid inequalities to asymmetric skill-based routing
  problems}.
\newblock {\it EURO Journal on Transportation and Logistics\/} {\bf 2}(1-2)
  29--55.

\bibitem[{Caramia and Guerriero(2009)}]{Caramia2009}
Caramia, M., F.~Guerriero. 2009.
\newblock {A heuristic approach to long-haul freight transportation with
  multiple objective functions}.
\newblock {\it Omega\/} {\bf 37} 600--614.

\bibitem[{Carlsson(2012)}]{Carlsson2012}
Carlsson, J.G. 2012.
\newblock {Dividing a territory among several vehicles}.
\newblock {\it INFORMS Journal on Computing\/} {\bf 24}(4) 565--577.

\bibitem[{Carlsson and Delage(2013)}]{Carlsson2013}
Carlsson, J.G., E.~Delage. 2013.
\newblock {Robust partitioning for stochastic multivehicle routing}.
\newblock {\it Operations Research\/} {\bf 61}(3) 727--744.

\bibitem[{Carlsson and Jia(2013)}]{Carlsson2013b}
Carlsson, J.G., F.~Jia. 2013.
\newblock {Euclidean hub-and-spoke networks}.
\newblock {\it Operations Research\/} {\bf 61}(6) 1360--1382.

\bibitem[{Carlsson and Jia(2015)}]{Carlsson2015}
Carlsson, J.G., F.~Jia. 2015.
\newblock {Continuous facility location with backbone network costs}.
\newblock {\it Transportation Science\/} {\bf 49}(3) 433--451.

\bibitem[{Cattaruzza et~al.(2016)Cattaruzza, Absi, and
  Feillet}]{Cattaruzza2016}
Cattaruzza, D., N.~Absi, D.~Feillet. 2016.
\newblock {The multi-trip vehicle routing problem with time windows and release
  dates}.
\newblock {\it Transportation Science\/} {\bf 50}(2) 676--693.

\bibitem[{{\c{C}}avdar and Sokol(2015)}]{Cavdar2015}
{\c{C}}avdar, B., J.~Sokol. 2015.
\newblock {A distribution-free TSP tour length estimation model for random
  graphs}.
\newblock {\it European Journal of Operational Research\/} {\bf 243}(2)
  588--598.

\bibitem[{{\c{C}}elik and S{\"{u}}ral(2016)}]{Celik2016}
{\c{C}}elik, M, H~S{\"{u}}ral. 2016.
\newblock {Order picking in a parallel-aisle warehouse with turn penalties}.
\newblock {\it International Journal of Production Research\/} {\bf 54}(14)
  1--16.

\bibitem[{Ceschia et~al.(2011)Ceschia, Gaspero, and Schaerf}]{Ceschia2011}
Ceschia, S., L.~Gaspero, A.~Schaerf. 2011.
\newblock {Tabu search techniques for the heterogeneous vehicle routing problem
  with time windows and carrier-dependent costs}.
\newblock {\it Journal of Scheduling\/} {\bf 14}(6) 601--615.

\bibitem[{Chandra(1993)}]{Chandra1993}
Chandra, P. 1993.
\newblock {A dynamic distribution model with warehouse and customer
  replenishment requirements}.
\newblock {\it Journal of the Operational Research Society\/} {\bf 44}(7)
  681--692.

\bibitem[{Chandra and Fisher(1994)}]{Chandra1994}
Chandra, P., M.L. Fisher. 1994.
\newblock {Coordination of production and distribution planning}.
\newblock {\it European Journal of Operational Research\/} {\bf 72} 503--517.

\bibitem[{Chemla et~al.(2013)Chemla, Meunier, and {Wolfler
  Calvo}}]{Chemla2013a}
Chemla, D., F.~Meunier, R.~{Wolfler Calvo}. 2013.
\newblock {Bike sharing systems: Solving the static rebalancing problem}.
\newblock {\it Discrete Optimization\/} {\bf 10}(2) 120--146.

\bibitem[{Chitsaz et~al.(2019)Chitsaz, Cordeau, and Jans}]{Chitsaz2019}
Chitsaz, M., J.-F. Cordeau, R.~Jans. 2019.
\newblock {A unified decomposition matheuristic for assembly, production, and
  inventory routing}.
\newblock {\it INFORMS Journal on Computing\/} {\bf 31}(1) 134--152.

\bibitem[{Christiansen et~al.(2013)Christiansen, Fagerholt, Nygreen, and
  Ronen}]{Christiansen2013}
Christiansen, M., K.~Fagerholt, B.~Nygreen, D.~Ronen. 2013.
\newblock {Ship routing and scheduling in the new millennium}.
\newblock {\it European Journal of Operational Research\/} {\bf 228}(3)
  467--483.

\bibitem[{Christofides(1971)}]{Christofides1971}
Christofides, N. 1971.
\newblock {Fixed routes and areas for delivery operations}.
\newblock {\it International Journal of Physical Distribution\/} {\bf 1}(2)
  87--92.

\bibitem[{Cleophas et~al.(2019)Cleophas, Cottrill, Ehmke, and
  Tierney}]{Cleophas2019}
Cleophas, C., C.~Cottrill, J.F. Ehmke, K.~Tierney. 2019.
\newblock {Collaborative urban transportation: Recent advances in theory and
  practice}.
\newblock {\it European Journal of Operational Research\/} {\bf 273}(3)
  801--816.

\bibitem[{Clossey et~al.(2001)Clossey, Laporte, and Soriano}]{Clossey2001}
Clossey, J., G.~Laporte, P.~Soriano. 2001.
\newblock {Solving arc routing problems with turn penalties}.
\newblock {\it Journal of the Operational Research Society\/} {\bf 52}(4)
  433--439.

\bibitem[{Coelho et~al.(2012)Coelho, Cordeau, and Laporte}]{Coelho2012a}
Coelho, L.C., J.-F. Cordeau, G.~Laporte. 2012.
\newblock {Consistency in multi-vehicle inventory-routing}.
\newblock {\it Transportation Research Part C: Emerging Technologies\/} {\bf
  24} 270--287.

\bibitem[{Coelho et~al.(2014)Coelho, Cordeau, and Laporte}]{Coelho2014}
Coelho, L.C., J.-F. Cordeau, G.~Laporte. 2014.
\newblock {Thirty years of inventory routing}.
\newblock {\it Transportation Science\/} {\bf 48}(1) 1--19.

\bibitem[{Coelho and Laporte(2013)}]{Coelho2013a}
Coelho, L.C., G.~Laporte. 2013.
\newblock {The exact solution of several classes of inventory-routing
  problems}.
\newblock {\it Computers {\&} Operations Research\/} {\bf 40}(2) 558--565.

\bibitem[{Coelho and Laporte(2014)}]{Coelho2014a}
Coelho, L.C., G.~Laporte. 2014.
\newblock {Optimal joint replenishment, delivery and inventory management
  policies for perishable products}.
\newblock {\it Computers {\&} Operations Research\/} {\bf 47} 42--52.

\bibitem[{Constantino et~al.(2015)Constantino, Gouveia, Mour{\~{a}}o, and
  Nunes}]{Constantino2015}
Constantino, M., L.~Gouveia, M.C. Mour{\~{a}}o, A.C. Nunes. 2015.
\newblock {The mixed capacitated arc routing problem with non-overlapping
  routes}.
\newblock {\it European Journal of Operational Research\/} {\bf 244}(2)
  445--456.

\bibitem[{Constantino et~al.(2017)Constantino, Mour{\~{a}}o, and
  Pinto}]{Constantino2017}
Constantino, M., M.C. Mour{\~{a}}o, L.S. Pinto. 2017.
\newblock {Dissimilar arc routing problems}.
\newblock {\it Networks\/} {\bf 70}(3) 233--245.

\bibitem[{Contardo et~al.(2014)Contardo, Cordeau, and Gendron}]{Contardo2014a}
Contardo, C., J.-F. Cordeau, B.~Gendron. 2014.
\newblock {An exact algorithm based on cut-and-column generation for the
  capacitated location-routing problem}.
\newblock {\it INFORMS Journal on Computing\/} {\bf 26}(1) 88--102.

\bibitem[{Cookson and Pishue(2018)}]{Cookson2018}
Cookson, G., B.~Pishue. 2018.
\newblock {INRIX Global Traffic Scorecard}.
\newblock Tech. rep., INRIX Research.

\bibitem[{Corber{\'{a}}n et~al.(2017)Corber{\'{a}}n, Golden, Lum, Plana, and
  Sanchis}]{Corberan2017}
Corber{\'{a}}n, {\'{A}}., B.L. Golden, O.~Lum, I.~Plana, J.M. Sanchis. 2017.
\newblock {Aesthetic considerations for the min-max $K$-windy rural postman
  problem}.
\newblock {\it Networks\/} {\bf 70}(3) 216--232.

\bibitem[{Corber{\'{a}}n et~al.(2002)Corber{\'{a}}n, Mart{\'{i}},
  Mart{\'{i}}nez, and Soler}]{Corberan2002}
Corber{\'{a}}n, {\'{A}}., R.~Mart{\'{i}}, E.~Mart{\'{i}}nez, D.~Soler. 2002.
\newblock {The rural postman problem on mixed graphs with turn penalties}.
\newblock {\it Journal of the Operational Research Society\/} {\bf 29}(7)
  887--903.

\bibitem[{Cordeau et~al.(2009)Cordeau, Iori, Laporte, and {Salazar
  Gonz{\'{a}}lez}}]{Cordeau2009}
Cordeau, J.-F., M.~Iori, G.~Laporte, J.-J. {Salazar Gonz{\'{a}}lez}. 2009.
\newblock {A branch-and-cut algorithm for the pickup and delivery traveling
  salesman problem with LIFO loading}.
\newblock {\it Networks\/} {\bf 55}(1) 46--59.

\bibitem[{Cordeau and Laporte(2007)}]{Cordeau2007a}
Cordeau, J.-F., G.~Laporte. 2007.
\newblock {The dial-a-ride problem: Models and algorithms}.
\newblock {\it Annals of Operations Research\/} {\bf 153}(1) 29--46.

\bibitem[{C{\^{o}}t{\'{e}} and Potvin(2009)}]{Cote2009a}
C{\^{o}}t{\'{e}}, J.-F., J.-Y. Potvin. 2009.
\newblock {A tabu search heuristic for the vehicle routing problem with private
  fleet and common carrier}.
\newblock {\it European Journal of Operational Research\/} {\bf 198}(2)
  464--469.

\bibitem[{Crainic(2002)}]{Crainic2002}
Crainic, T.G. 2002.
\newblock {Long-haul freight transportation}.
\newblock R.W. Hall, ed., {\it Handbook of Transportation Science\/}. Springer,
  Boston, 451--516.

\bibitem[{Crama et~al.(2018)Crama, Rezaei, Savelsbergh, and {Van
  Woensel}}]{Crama2018}
Crama, Y., M.~Rezaei, M.W.P. Savelsbergh, T.~{Van Woensel}. 2018.
\newblock {Stochastic inventory routing for perishable products}.
\newblock {\it Transportation Science\/} {\bf 52}(3) 526--546.

\bibitem[{Cruijssen et~al.(2007)Cruijssen, Dullaert, and
  Fleuren}]{Cruijssen2007a}
Cruijssen, F., W.~Dullaert, H.~Fleuren. 2007.
\newblock {Horizontal cooperation in transport and logistics: A literature
  review}.
\newblock {\it Transportation Journal\/} {\bf 46}(3) 22--39.

\bibitem[{Cuda et~al.(2015)Cuda, Guastaroba, and Speranza}]{Cuda2015}
Cuda, R., G.~Guastaroba, M.G. Speranza. 2015.
\newblock {A survey on two-echelon routing problems}.
\newblock {\it Computers {\&} Operations Research\/} {\bf 55} 185--199.

\bibitem[{Dabia et~al.(2019)Dabia, Lai, and Vigo}]{Dabia2019}
Dabia, S., D.~Lai, D.~Vigo. 2019.
\newblock {An exact algorithm for a rich vehicle routing problem with private
  fleet and common carrier}.
\newblock {\it Transportation Science\/} {\bf 53}(4) 986--1000.

\bibitem[{Daganzo(1984)}]{Daganzo1984}
Daganzo, C.F. 1984.
\newblock {The distance traveled to visit N points with a maximum of C stops
  per vehicle: An analytic model and an application}.
\newblock {\it Transportation Science\/} {\bf 18}(4) 331--350.

\bibitem[{Darvish and Coelho(2018)}]{Darvish2018}
Darvish, M., L.C. Coelho. 2018.
\newblock {Sequential versus integrated optimization: Production, location,
  inventory control, and distribution}.
\newblock {\it European Journal of Operational Research\/} {\bf 268}(1)
  203--214.

\bibitem[{Davis and Figliozzi(2013)}]{Davis2013}
Davis, B.A., M.A. Figliozzi. 2013.
\newblock {A methodology to evaluate the competitiveness of electric delivery
  trucks}.
\newblock {\it Transportation Research Part E: Logistics and Transportation
  Review\/} {\bf 49}(1) 8--23.

\bibitem[{Day et~al.(2009)Day, Wright, Schoenherr, Venkataramanan, and
  Gaudette}]{Day2009}
Day, J.M., P.D. Wright, T.~Schoenherr, M.~Venkataramanan, K.~Gaudette. 2009.
\newblock {Improving routing and scheduling decisions at a distributor of
  industrial gasses}.
\newblock {\it Omega\/} {\bf 37}(1) 227--237.

\bibitem[{Dell'Amico et~al.(2015)Dell'Amico, Falavigna, and
  Iori}]{DellAmico2015}
Dell'Amico, M., S.~Falavigna, M.~Iori. 2015.
\newblock {Optimization of a real-world auto-carrier transportation problem}.
\newblock {\it Transportation Science\/} {\bf 49}(2) 402--419.

\bibitem[{Demir et~al.(2011)Demir, Bektaş, and Laporte}]{Demir2011}
Demir, E., T.~Bektaş, G.~Laporte. 2011.
\newblock {A comparative analysis of several vehicle emission models for road
  freight transportation}.
\newblock {\it Transportation Research Part D: Transport and Environment\/}
  {\bf 16}(5) 347--357.

\bibitem[{Demir et~al.(2014{\natexlab{a}})Demir, Bektaş, and
  Laporte}]{Demir2014a}
Demir, E., T.~Bektaş, G.~Laporte. 2014{\natexlab{a}}.
\newblock {A review of recent research on green road freight transportation}.
\newblock {\it European Journal of Operational Research\/} {\bf 237}(3)
  775--793.

\bibitem[{Demir et~al.(2014{\natexlab{b}})Demir, Bektaş, and
  Laporte}]{Demir2014}
Demir, E., T.~Bektaş, G.~Laporte. 2014{\natexlab{b}}.
\newblock {The bi-objective pollution-routing problem}.
\newblock {\it European Journal of Operational Research\/} {\bf 232}(3)
  464--478.

\bibitem[{Demir et~al.(2015)Demir, Huang, Scholts, and {Van
  Woensel}}]{Demir2015}
Demir, E., Y.~Huang, S.~Scholts, T.~{Van Woensel}. 2015.
\newblock {A selected review on the negative externalities of the freight
  transportation: Modeling and pricing}.
\newblock {\it Transportation Research Part E: Logistics and Transportation
  Review\/} {\bf 77} 95--114.

\bibitem[{Derigs et~al.(2010)Derigs, Gottlieb, Kalkoff, Piesche, Rothlauf, and
  Vogel}]{Derigs2010}
Derigs, U., J.~Gottlieb, J.~Kalkoff, M.~Piesche, F.~Rothlauf, U.~Vogel. 2010.
\newblock {Vehicle routing with compartments: Applications, modelling and
  heuristics}.
\newblock {\it OR Spectrum\/} {\bf 33}(4) 885--914.

\bibitem[{Desaulniers et~al.(2016)Desaulniers, Rakke, and
  Coelho}]{Desaulniers2016a}
Desaulniers, G., J.G. Rakke, L.C. Coelho. 2016.
\newblock {A branch-price-and-cut algorithm for the inventory-routing problem}.
\newblock {\it Transportation Science\/} {\bf 50}(3) 1060--1076.

\bibitem[{Doerner and Salazar-Gonz{\'{a}}lez(2014)}]{Doerner2014}
Doerner, K.F., J.-J. Salazar-Gonz{\'{a}}lez. 2014.
\newblock {Pickup-and-delivery problems for people transportation}.
\newblock P.~Toth, D.~Vigo, eds., {\it Vehicle Routing: Problems, Methods, and
  Applications\/}. SIAM, Philadelphia, 193--212.

\bibitem[{Dorling et~al.(2017)Dorling, Heinrichs, Messier, and
  Magierowski}]{Dorling2017}
Dorling, K., J.~Heinrichs, G.G. Messier, S.~Magierowski. 2017.
\newblock {Vehicle routing problems for drone delivery}.
\newblock {\it IEEE Transactions on Systems, Man, and Cybernetics: Systems\/}
  {\bf 47}(1) 70--85.

\bibitem[{Drexl and Haase(1999)}]{Drexl1999}
Drexl, A., K.~Haase. 1999.
\newblock {Fast approximation methods for sales force deployment}.
\newblock {\it Management Science\/} {\bf 45}(10) 1307--1323.

\bibitem[{Drexl(2012)}]{Drexl2012a}
Drexl, M. 2012.
\newblock {Synchronization in vehicle routing -- A survey of VRPs with multiple
  synchronization constraints}.
\newblock {\it Transportation Science\/} {\bf 46}(3) 297--316.

\bibitem[{Dror et~al.(1989)Dror, Laporte, and Trudeau}]{Dror1989}
Dror, M., G.~Laporte, P.~Trudeau. 1989.
\newblock {Vehicle routing with stochastic demands: Properties and solution
  frameworks}.
\newblock {\it Transportation Science\/} {\bf 23}(3) 166--176.

\bibitem[{Duckham and Kulik(2003)}]{Duckham2003}
Duckham, M., L.~Kulik. 2003.
\newblock {Simplest paths: Automated route selection for navigation}.
\newblock W.~Kuhn, M.F. Worboys, S.~Timpf, eds., {\it International Conference
  on Spatial Information Theory\/}. Springer, Kartause Ittingen, Switzerland,
  169--185.

\bibitem[{Errico et~al.(2018)Errico, Desaulniers, Gendreau, Rei, and
  Rousseau}]{Errico2018}
Errico, F., G.~Desaulniers, M.~Gendreau, W.~Rei, L.-M. Rousseau. 2018.
\newblock {The vehicle routing problem with hard time windows and stochastic
  service times}.
\newblock {\it EURO Journal on Transportation and Logistics\/} {\bf 7}(3)
  223--251.

\bibitem[{Eveborn et~al.(2006)Eveborn, Flisberg, and
  R{\"{o}}nnqvist}]{Eveborn2006}
Eveborn, P., P.~Flisberg, M.~R{\"{o}}nnqvist. 2006.
\newblock {LAPS CARE -- An operational system for staff planning of home care}.
\newblock {\it European Journal of Operational Research\/} {\bf 171}(1)
  962--976.

\bibitem[{Eveborn et~al.(2009)Eveborn, R{\"{o}}nnqvist, Einarsd{\'{o}}ttir,
  Eklund, Lid{\'{e}}n, and Almroth}]{Eveborn2009}
Eveborn, P., M.~R{\"{o}}nnqvist, H.~Einarsd{\'{o}}ttir, M.~Eklund,
  K.~Lid{\'{e}}n, M.~Almroth. 2009.
\newblock {Operations research improves quality and efficiency in home care}.
\newblock {\it Interfaces\/} {\bf 39}(1) 18--34.

\bibitem[{Fagerholt et~al.(2009)Fagerholt, Laporte, and
  Norstad}]{Fagerholt2009}
Fagerholt, K., G.~Laporte, I.~Norstad. 2009.
\newblock {Reducing fuel emissions by optimizing speed on shipping routes}.
\newblock {\it Journal of the Operational Research Society\/} {\bf 61}(3)
  523--529.

\bibitem[{Fagnant and Kockelman(2015)}]{Fagnant2015}
Fagnant, D.J., K.~Kockelman. 2015.
\newblock {Preparing a nation for autonomous vehicles: Opportunities, barriers
  and policy recommendations}.
\newblock {\it Transportation Research Part A: Policy and Practice\/} {\bf 77}
  167--181.

\bibitem[{Feillet et~al.(2014)Feillet, Garaix, Lehuede, P{\'{e}}ton, and
  Quadri}]{Feillet2014}
Feillet, D., T.~Garaix, F.~Lehuede, O.~P{\'{e}}ton, D.~Quadri. 2014.
\newblock {A new consistent vehicle routing problem for the transportation of
  people with disabilities}.
\newblock {\it Networks\/} {\bf 63}(3) 211--224.

\bibitem[{Felipe et~al.(2014)Felipe, Ortu{\~{n}}o, Righini, and
  Tirado}]{Felipe2014}
Felipe, {\'{A}}., M.T. Ortu{\~{n}}o, G.~Righini, G.~Tirado. 2014.
\newblock {A heuristic approach for the green vehicle routing problem with
  multiple technologies and partial recharges}.
\newblock {\it Transportation Research Part E: Logistics and Transportation
  Review\/} {\bf 71} 111--128.

\bibitem[{Figliozzi and Tipagornwong(2017)}]{Figliozzi2017}
Figliozzi, M., C.~Tipagornwong. 2017.
\newblock {Impact of last mile parking availability on commercial vehicle costs
  and operations}.
\newblock {\it Supply Chain Forum: An International Journal\/} {\bf 18}(2)
  60--68.

\bibitem[{Franceschetti et~al.(2017{\natexlab{a}})Franceschetti, Honhon,
  Laporte, {Van Woensel}, and Fransoo}]{Franceschetti2017a}
Franceschetti, A., D.~Honhon, G.~Laporte, T.~{Van Woensel}, J.C. Fransoo.
  2017{\natexlab{a}}.
\newblock {Strategic fleet planning for city logistics}.
\newblock {\it Transportation Research Part B: Methodological\/} {\bf 95}
  19--40.

\bibitem[{Franceschetti et~al.(2013)Franceschetti, Honhon, {Van Woensel},
  Bektaş, and Laporte}]{Franceschetti2013}
Franceschetti, A., D.~Honhon, T.~{Van Woensel}, T.~Bektaş, G.~Laporte. 2013.
\newblock {The time-dependent pollution-routing problem}.
\newblock {\it Transportation Research Part B: Methodological\/} {\bf 56}
  265--293.

\bibitem[{Franceschetti et~al.(2017{\natexlab{b}})Franceschetti, Jabali, and
  Laporte}]{Franceschetti2017}
Franceschetti, A., O.~Jabali, G.~Laporte. 2017{\natexlab{b}}.
\newblock {Continuous approximation models in freight distribution management}.
\newblock {\it TOP\/} {\bf 25}(3) 413--433.

\bibitem[{Froger et~al.(2017)Froger, Mendoza, Jabali, and Laporte}]{Froger2017}
Froger, A., J.E. Mendoza, O.~Jabali, G.~Laporte. 2017.
\newblock {A matheuristic for the electric vehicle routing problem with
  capacitated charging stations}.
\newblock Tech. rep., CIRRELT-2017-31.

\bibitem[{Fukasawa et~al.(2018)Fukasawa, He, Santos, and Song}]{Fukasawa2018}
Fukasawa, R., Q.~He, F.~Santos, Y.~Song. 2018.
\newblock {A joint vehicle routing and speed optimization problem}.
\newblock {\it INFORMS Journal on Computing\/} {\bf 30}(4) 694--709.

\bibitem[{Gahm et~al.(2017)Gahm, Brab{\"{a}}nder, and Tuma}]{Gahm2017}
Gahm, C., C.~Brab{\"{a}}nder, A.~Tuma. 2017.
\newblock {Vehicle routing with private fleet, multiple common carriers
  offering volume discounts, and rental options}.
\newblock {\it Transportation Research Part E: Logistics and Transportation
  Review\/} {\bf 97} 192--216.

\bibitem[{Gansterer and Hartl(2018)}]{Gansterer2018}
Gansterer, M., R.F. Hartl. 2018.
\newblock {Collaborative vehicle routing: A survey}.
\newblock {\it European Journal of Operational Research\/} {\bf 268}(1) 1--12.

\bibitem[{Garaix et~al.(2010)Garaix, Artigues, Feillet, and
  Josselin}]{Garaix2010a}
Garaix, T., C.~Artigues, D.~Feillet, D.~Josselin. 2010.
\newblock {Vehicle routing problems with alternative paths: An application to
  on-demand transportation}.
\newblock {\it European Journal of Operational Research\/} {\bf 204} 62--75.

\bibitem[{Garaix et~al.(2011)Garaix, Artigues, Feillet, and
  Josselin}]{Garaix2011}
Garaix, T., C.~Artigues, D.~Feillet, D.~Josselin. 2011.
\newblock {Optimization of occupancy rate in dial-a-ride problems via linear
  fractional column generation}.
\newblock {\it Computers {\&} Operations Research\/} {\bf 38}(10) 1435--1442.

\bibitem[{Gaudioso and Paletta(1992)}]{Gaudioso1992}
Gaudioso, M., G.~Paletta. 1992.
\newblock {A heuristic for the periodic vehicle routing problem}.
\newblock {\it Transportation Science\/} {\bf 26}(2) 86--92.

\bibitem[{Gendreau et~al.(2015)Gendreau, Ghiani, and Guerriero}]{Gendreau2015}
Gendreau, M., G.~Ghiani, E.~Guerriero. 2015.
\newblock {Time-dependent routing problems: A review}.
\newblock {\it Computers {\&} Operations Research\/} {\bf 64} 189--197.

\bibitem[{Gendreau et~al.(2006)Gendreau, Iori, Laporte, and
  Martello}]{Gendreau2006a}
Gendreau, M., M.~Iori, G.~Laporte, S.~Martello. 2006.
\newblock {A tabu search algorithm for a routing and container loading
  problem}.
\newblock {\it Transportation Science\/} {\bf 40}(3) 342--350.

\bibitem[{Gendreau et~al.(2014)Gendreau, Jabali, and Rei}]{Gendreau2014}
Gendreau, M., O.~Jabali, W.~Rei. 2014.
\newblock {Stochastic vehicle routing problems}.
\newblock P.~Toth, D.~Vigo, eds., {\it Vehicle Routing: Problems, Methods, and
  Applications\/}. Society for Industrial and Applied Mathematics,
  Philadelphia.

\bibitem[{Gendreau et~al.(2016)Gendreau, Jabali, and Rei}]{Gendreau2016}
Gendreau, M., O.~Jabali, W.~Rei. 2016.
\newblock {Future research directions in stochastic vehicle routing}.
\newblock {\it Transportation Science\/} {\bf 50}(4) 1163--1173.

\bibitem[{Ghiami et~al.(2019)Ghiami, Demir, {Van Woensel}, Christiansen, and
  Laporte}]{Ghiami2019}
Ghiami, Y., E.~Demir, T.~{Van Woensel}, M.~Christiansen, G.~Laporte. 2019.
\newblock {A deteriorating inventory routing problem for an inland liquefied
  natural gas distribution network}.
\newblock {\it Transportation Research Part B: Methodological\/} {\bf 126}
  45--67.

\bibitem[{Goeke et~al.(2019{\natexlab{a}})Goeke, Gschwind, and
  Schneider}]{Goeke2019a}
Goeke, D., T.~Gschwind, M.~Schneider. 2019{\natexlab{a}}.
\newblock {Upper and lower bounds for the vehicle-routing problem with private
  fleet and common carrier}.
\newblock {\it Discrete Applied Mathematics\/} {\bf 264} 43--61.

\bibitem[{Goeke et~al.(2019{\natexlab{b}})Goeke, Roberti, and
  Schneider}]{Goeke2019}
Goeke, D., R.~Roberti, M.~Schneider. 2019{\natexlab{b}}.
\newblock {Exact and heuristic solution of the consistent vehicle-routing
  problem}.
\newblock {\it Transportation Science\/} {\bf 53}(4) 1023--1042.

\bibitem[{Goeke and Schneider(2015)}]{Goeke2015}
Goeke, D., M.~Schneider. 2015.
\newblock {Routing a mixed fleet of electric and conventional vehicles}.
\newblock {\it European Journal of Operational Research\/} {\bf 245}(1) 81--99.

\bibitem[{Goel(2018)}]{Goel2018}
Goel, A. 2018.
\newblock {Legal aspects in road transport optimization in Europe}.
\newblock {\it Transportation Research Part E: Logistics and Transportation
  Review\/} {\bf 114} 144--162.

\bibitem[{Goel and Meisel(2013)}]{Goel2013}
Goel, A., F.~Meisel. 2013.
\newblock {Workforce routing and scheduling for electricity network maintenance
  with downtime minimization}.
\newblock {\it European Journal of Operational Research\/} {\bf 231}(1)
  210--228.

\bibitem[{Goel and Vidal(2014)}]{Goel2012}
Goel, A., T.~Vidal. 2014.
\newblock {Hours of service regulations in road freight transport: An
  optimization-based international assessment}.
\newblock {\it Transportation Science\/} {\bf 48}(3) 391--412.

\bibitem[{Goel et~al.(2019)Goel, Vidal, and Kok}]{Goel2019}
Goel, A., T.~Vidal, A.L. Kok. 2019.
\newblock {To team up or not – Single versus team driving in European road
  freight transport}.
\newblock Tech. rep., PUC--Rio, Rio de Janeiro, Brasil.

\bibitem[{Golden et~al.(2008)Golden, Raghavan, and Wasil}]{Golden2008}
Golden, B., S.~Raghavan, E.~Wasil, eds. 2008.
\newblock {\it {The Vehicle Routing Problem: Latest Advances and New
  Challenges}\/}.
\newblock Springer, New York.

\bibitem[{Golden et~al.(2014)Golden, Kovacs, and Wasil}]{Golden2014}
Golden, B.L., A.A. Kovacs, E.A. Wasil. 2014.
\newblock {Vehicle routing applications in disaster relief}.
\newblock P.~Toth, D.~Vigo, eds., {\it Vehicle Routing: Problems, Methods, and
  Applications\/}. Society for Industrial and Applied Mathematics,
  Philadelphia, 409--436.

\bibitem[{Goodson(2014)}]{Goodson2014}
Goodson, J.C. 2014.
\newblock {Election day routing of rapid response attorneys}.
\newblock {\it INFOR: Information Systems and Operational Research\/} {\bf
  52}(1) 1--9.

\bibitem[{Goodson et~al.(2013)Goodson, Ohlmann, and Thomas}]{Goodson2013}
Goodson, J.C., J.W. Ohlmann, B.W. Thomas. 2013.
\newblock {Rollout policies for dynamic solutions to the multivehicle routing
  problem with stochastic demand and duration limits}.
\newblock {\it Operations Research\/} {\bf 61}(1) 138--154.

\bibitem[{Gounaris et~al.(2013)Gounaris, Wiesemann, and Floudas}]{Gounaris2013}
Gounaris, C.E., W.~Wiesemann, C.A. Floudas. 2013.
\newblock {The robust capacitated vehicle routing problem under demand
  uncertainty}.
\newblock {\it Operations Research\/} {\bf 61}(3) 677--693.

\bibitem[{Grabenschweiger et~al.(2018)Grabenschweiger, Tricoire, and
  Doerner}]{Grabenschweiger2018}
Grabenschweiger, J., F.~Tricoire, K.F. Doerner. 2018.
\newblock {Finding the trade-off between emissions and disturbance in an urban
  context}.
\newblock {\it Flexible Services and Manufacturing Journal\/} {\bf 30}(3)
  554--591.

\bibitem[{Grangier et~al.(2016)Grangier, Gendreau, Lehu{\'{e}}d{\'{e}}, and
  Rousseau}]{Grangier2016}
Grangier, P., M.~Gendreau, F.~Lehu{\'{e}}d{\'{e}}, L.-M. Rousseau. 2016.
\newblock {An adaptive large neighborhood search for the two-echelon
  multiple-trip vehicle routing problem with satellite synchronization}.
\newblock {\it European Journal of Operational Research\/} {\bf 254}(1) 80--91.

\bibitem[{Gro{\"{e}}r and Golden(2009)}]{Groer2009}
Gro{\"{e}}r, C., B.L. Golden. 2009.
\newblock {The balanced billing cycle vehicle routing problem}.
\newblock {\it Networks\/} {\bf 54}(4) 243--254.

\bibitem[{Gschwind and Drexl(2019)}]{Gschwind2019}
Gschwind, T., M.~Drexl. 2019.
\newblock {Adaptive large neighborhood search with a constant-time feasibility
  test for the dial-a-ride problem}.
\newblock {\it Transportation Science\/} {\bf 53}(2) 480--491.

\bibitem[{Guajardo and R{\"{o}}nnqvist(2016)}]{Guajardo2016}
Guajardo, M., M.~R{\"{o}}nnqvist. 2016.
\newblock {A review on cost allocation methods in collaborative
  transportation}.
\newblock {\it International Transactions in Operational Research\/} {\bf
  23}(3) 371--392.

\bibitem[{Guastaroba et~al.(2016)Guastaroba, Speranza, and
  Vigo}]{Guastaroba2016}
Guastaroba, G., M.G. Speranza, D.~Vigo. 2016.
\newblock {Intermediate facilities in freight transportation planning: A
  survey}.
\newblock {\it Transportation Science\/} {\bf 50}(3) 763--789.

\bibitem[{Haddad et~al.(2018)Haddad, Martinelli, Vidal, Martins, Ochi, Souza,
  and Hartl}]{Haddad2018}
Haddad, M.N., R.~Martinelli, T.~Vidal, S.~Martins, L.S. Ochi, M.J.F. Souza,
  R.F. Hartl. 2018.
\newblock {Large neighborhood-based metaheuristic and branch-and-price for the
  pickup and delivery problem with split loads}.
\newblock {\it European Journal of Operational Research\/} {\bf 270}(3)
  1014--1027.

\bibitem[{Hall and Partyka(2018)}]{Hall2018}
Hall, R., J.~Partyka. 2018.
\newblock {Higher expectations drive transformation}.
\newblock {\it OR/MS Today\/} {\bf 45}(1) 40--47.

\bibitem[{Halvorsen-Weare and Fagerholt(2013)}]{HalvorsenWeare2013}
Halvorsen-Weare, E.E., K.~Fagerholt. 2013.
\newblock {Routing and scheduling in a liquefied natural gas shipping problem
  with inventory and berth constraints}.
\newblock {\it Annals of Operations Research\/} {\bf 203}(1) 167--186.

\bibitem[{Hamdi-Dhaoui et~al.(2014)Hamdi-Dhaoui, Labadie, and
  Yalaoui}]{Hamdi-Dhaoui2014}
Hamdi-Dhaoui, K., N.~Labadie, A.~Yalaoui. 2014.
\newblock {The bi-objective two-dimensional loading vehicle routing problem
  with partial conflicts}.
\newblock {\it International Journal of Production Research\/} {\bf 52}(19)
  5565--5582.

\bibitem[{Hartl and Romauch(2016)}]{Hartl2015}
Hartl, R.F., M.~Romauch. 2016.
\newblock {Notes on the single route lateral transhipment problem}.
\newblock {\it Journal of Global Optimization\/} {\bf 65}(1) 57--82.

\bibitem[{Haughton(2002)}]{Haughton2002}
Haughton, M.A. 2002.
\newblock {Measuring and managing the learning requirements of route
  reoptimization on delivery vehicle drivers}.
\newblock {\it Journal of Business Logistics\/} {\bf 23}(2) 45--66.

\bibitem[{Haughton(2007)}]{Haughton2007}
Haughton, M.A. 2007.
\newblock {Assigning delivery routes to drivers under variable customer
  demands}.
\newblock {\it Transportation Research Part E: Logistics and Transportation
  Review\/} {\bf 43}(1) 157--172.

\bibitem[{Hiermann et~al.(2019{\natexlab{a}})Hiermann, Hartl, Puchinger,
  Schiffer, and Vidal}]{Hiermann2019a}
Hiermann, G., R.F. Hartl, J.~Puchinger, M.~Schiffer, T.~Vidal.
  2019{\natexlab{a}}.
\newblock {Sustainable city logistics via access restrictions? An impact
  assessment of city center policies}.
\newblock Tech. rep., University of Vienna, Austria.

\bibitem[{Hiermann et~al.(2019{\natexlab{b}})Hiermann, Hartl, Puchinger, and
  Vidal}]{Hiermann2019}
Hiermann, G., R.F. Hartl, J.~Puchinger, T.~Vidal. 2019{\natexlab{b}}.
\newblock {Routing a mix of conventional, plug-in hybrid, and electric
  vehicles}.
\newblock {\it European Journal of Operational Research\/} {\bf 272}(1)
  235--248.

\bibitem[{Hiermann et~al.(2016)Hiermann, Puchinger, Ropke, and
  Hartl}]{Hiermann2016}
Hiermann, G., J.~Puchinger, S.~Ropke, R.F. Hartl. 2016.
\newblock {The electric fleet size and mix vehicle routing problem with time
  windows and recharging stations}.
\newblock {\it European Journal of Operational Research\/} {\bf 252}(3)
  995--1018.

\bibitem[{Hoff et~al.(2010)Hoff, Andersson, Christiansen, Hasle, and
  L{\o}kketangen}]{Hoff2010}
Hoff, A., H.~Andersson, M.~Christiansen, G.~Hasle, A.~L{\o}kketangen. 2010.
\newblock {Industrial aspects and literature survey: Fleet composition and
  routing}.
\newblock {\it Computers {\&} Operations Research\/} {\bf 37}(12) 2041--2061.

\bibitem[{Holland et~al.(2017)Holland, Levis, Nuggehalli, Santilli, and
  Winters}]{Holland2017}
Holland, C., J.~Levis, R.~Nuggehalli, B.~Santilli, J.~Winters. 2017.
\newblock {UPS optimizes delivery routes}.
\newblock {\it Interfaces\/} {\bf 47}(1) 8--23.

\bibitem[{Hollis and Green(2012)}]{Hollis2012}
Hollis, B.L., P.J. Green. 2012.
\newblock {Real-life vehicle routing with time windows for visual
  attractiveness and operational robustness}.
\newblock {\it Asia-Pacific Journal of Operational Research\/} {\bf 29}(4)
  1--29.

\bibitem[{Hoogeboom and Dullaert(2019)}]{Hoogeboom2019}
Hoogeboom, M., W.~Dullaert. 2019.
\newblock {Vehicle routing with arrival time diversification}.
\newblock {\it European Journal of Operational Research\/} {\bf 275}(1)
  93--107.

\bibitem[{Huang et~al.(2012)Huang, Smilowitz, and Balcik}]{Huang2012}
Huang, M., K.~Smilowitz, B.~Balcik. 2012.
\newblock {Models for relief routing: Equity, efficiency and efficacy}.
\newblock {\it Transportation Research Part E: Logistics and Transportation
  Review\/} {\bf 48}(1) 2--18.

\bibitem[{Huang et~al.(2017)Huang, Zhao, {Van Woensel}, and Gross}]{Huang2017}
Huang, Y., L.~Zhao, T.~{Van Woensel}, J.-P. Gross. 2017.
\newblock {Time-dependent vehicle routing problem with path flexibility}.
\newblock {\it Transportation Research Part B: Methodological\/} {\bf 95}
  169--195.

\bibitem[{Hvattum et~al.(2013)Hvattum, Norstad, Fagerholt, and
  Laporte}]{Hvattum2013}
Hvattum, L.M., I.~Norstad, K.~Fagerholt, G.~Laporte. 2013.
\newblock {Analysis of an exact algorithm for the vessel speed optimization
  problem}.
\newblock {\it Networks\/} {\bf 62}(2) 132--135.

\bibitem[{Hwang et~al.(2017)Hwang, Myung, and Sun}]{Hwang2017}
Hwang, J.-Y., S.-T. Myung, Y.-K. Sun. 2017.
\newblock {Sodium-ion batteries: Present and future}.
\newblock {\it Chemical Society Reviews\/} {\bf 46}(12) 3485--3856.

\bibitem[{Ichoua et~al.(2003)Ichoua, Gendreau, and Potvin}]{Ichoua2003}
Ichoua, S., M.~Gendreau, J.-Y. Potvin. 2003.
\newblock {Vehicle dispatching with time-dependent travel times}.
\newblock {\it European Journal of Operational Research\/} {\bf 144}(2)
  379--396.

\bibitem[{Iori et~al.(2007)Iori, {Salazar Gonz{\'{a}}lez}, and Vigo}]{Iori2007}
Iori, M., J.J. {Salazar Gonz{\'{a}}lez}, D.~Vigo. 2007.
\newblock {An exact approach for the vehicle routing problem with
  two-dimensional loading constraints}.
\newblock {\it Transportation Science\/} {\bf 41}(2) 253--264.

\bibitem[{Irnich(2008)}]{Irnich2008}
Irnich, S. 2008.
\newblock {Solution of real-world postman problems}.
\newblock {\it European Journal of Operational Research\/} {\bf 190}(1) 52--67.

\bibitem[{Jabali et~al.(2012{\natexlab{a}})Jabali, Gendreau, and
  Laporte}]{Jabali2012a}
Jabali, O., M.~Gendreau, G.~Laporte. 2012{\natexlab{a}}.
\newblock {A continuous approximation model for the fleet composition problem}.
\newblock {\it Transportation Research Part B: Methodological\/} {\bf 46}(10)
  1591--1606.

\bibitem[{Jabali et~al.(2015)Jabali, Leus, {Van Woensel}, and
  de~Kok}]{Jabali2015}
Jabali, O., R.~Leus, T.~{Van Woensel}, T.~de~Kok. 2015.
\newblock {Self-imposed time windows in vehicle routing problems}.
\newblock {\it OR Spectrum\/} {\bf 37}(2) 331--352.

\bibitem[{Jabali et~al.(2012{\natexlab{b}})Jabali, {Van Woensel}, and
  de~Kok}]{Jabali2012}
Jabali, O., T.~{Van Woensel}, A.G. de~Kok. 2012{\natexlab{b}}.
\newblock {Analysis of travel times and CO2 emissions in time-dependent vehicle
  routing}.
\newblock {\it Production and Operations Management\/} {\bf 21}(6) 1060--1074.

\bibitem[{Jacobsen and Madsen(1980)}]{Jacobsen1980}
Jacobsen, S.K., O.B.G. Madsen. 1980.
\newblock {A comparative study of heuristics for a two-level routing-location
  problem}.
\newblock {\it European Journal of Operational Research\/} {\bf 5}(6) 378--387.

\bibitem[{Jaillet(1988)}]{Jaillet1988b}
Jaillet, P. 1988.
\newblock {A priori solution of a traveling salesman problem in which a random
  subset of the customers are visited}.
\newblock {\it Operations Research\/} {\bf 36}(6) 929--936.

\bibitem[{Jaillet et~al.(2016)Jaillet, Qi, and Sim}]{Jaillet2016a}
Jaillet, P., J.~Qi, M.~Sim. 2016.
\newblock {Routing optimization under uncertainty}.
\newblock {\it Operations Research\/} {\bf 64}(1) 186--200.

\bibitem[{Jain(2010)}]{Jain2010}
Jain, A.K. 2010.
\newblock {Data clustering: 50 years beyond $K$-means}.
\newblock {\it Pattern Recognition Letters\/} {\bf 31}(8) 651--666.

\bibitem[{Janssens et~al.(2015)Janssens, {Van den Bergh}, S{\"{o}}rensen, and
  Cattrysse}]{Janssens2015}
Janssens, J., J.~{Van den Bergh}, K.~S{\"{o}}rensen, D.~Cattrysse. 2015.
\newblock {Multi-objective microzone-based vehicle routing for courier
  companies: From tactical to operational planning}.
\newblock {\it European Journal of Operational Research\/} {\bf 242}(1)
  222--231.

\bibitem[{Kalcsics(2015)}]{Kalcsics2015}
Kalcsics, J. 2015.
\newblock {Districting problems}.
\newblock G.~Laporte, S.~Nickel, F.~{Saldanha da Gama}, eds., {\it Location
  Science\/}. Springer, Berlin, 595--622.

\bibitem[{Kara et~al.(2007)Kara, Kara, and Yetis}]{Kara2007}
Kara, I., B.Y. Kara, M.K. Yetis. 2007.
\newblock {Energy minimizing vehicle routing problem}.
\newblock {\it International Conference on Combinatorial Optimization and
  Applications\/}. Springer, Berlin Heidelberg, 62--71.

\bibitem[{Keskin and {\c{C}}atay(2016)}]{Keskin2016}
Keskin, M., B.~{\c{C}}atay. 2016.
\newblock {Partial recharge strategies for the electric vehicle routing problem
  with time windows}.
\newblock {\it Transportation Research Part C: Emerging Technologies\/} {\bf
  65} 111--127.

\bibitem[{Kilby and Urli(2016)}]{Kilby2016}
Kilby, P., T.~Urli. 2016.
\newblock {Fleet design optimisation from historical data using constraint
  programming and large neighbourhood search}.
\newblock {\it Constraints\/} {\bf 21} 2--21.

\bibitem[{Kim et~al.(2014)Kim, Cha, and Sandholm}]{Kim2014}
Kim, J., M.~Cha, T.~Sandholm. 2014.
\newblock {Socroutes: Safe routes based on tweet sentiments}.
\newblock {\it Proceedings of the 23rd International Conference on World Wide
  Web\/}. Association for Computing Machinery, 179--182.

\bibitem[{Klibi et~al.(2010)Klibi, Lasalle, Martel, and Ichoua}]{Klibi2010}
Klibi, W., F.~Lasalle, A.~Martel, S.~Ichoua. 2010.
\newblock {The stochastic multiperiod location transportation problem}.
\newblock {\it Transportation Science\/} {\bf 44}(2) 221--237.

\bibitem[{Ko{\c{c}} et~al.(2015)Ko{\c{c}}, Bektaş, Jabali, and
  Laporte}]{Koc2015}
Ko{\c{c}}, {\c{C}}., T.~Bektaş, O.~Jabali, G.~Laporte. 2015.
\newblock {A hybrid evolutionary algorithm for heterogeneous fleet vehicle
  routing problems with time windows}.
\newblock {\it Computers {\&} Operations Research\/} {\bf 64} 11--27.

\bibitem[{Ko{\c{c}} et~al.(2016)Ko{\c{c}}, Bektaş, Jabali, and
  Laporte}]{Koc2016}
Ko{\c{c}}, {\c{C}}., T.~Bektaş, O.~Jabali, G.~Laporte. 2016.
\newblock {Thirty years of heterogeneous vehicle routing}.
\newblock {\it European Journal of Operational Research\/} {\bf 249}(1) 1--21.

\bibitem[{Kovacs et~al.(2014{\natexlab{a}})Kovacs, Golden, Hartl, and
  Parragh}]{Kovacs2014}
Kovacs, A.A., B.L. Golden, R.F. Hartl, S.N. Parragh. 2014{\natexlab{a}}.
\newblock {Vehicle routing problems in which consistency considerations are
  important: A survey}.
\newblock {\it Networks\/} {\bf 64}(3) 192--213.

\bibitem[{Kovacs et~al.(2014{\natexlab{b}})Kovacs, Parragh, and
  Hartl}]{Kovacs2014b}
Kovacs, A.A., S.N. Parragh, R.F. Hartl. 2014{\natexlab{b}}.
\newblock {A template-based adaptive large neighborhood search for the
  consistent vehicle routing problem}.
\newblock {\it Networks\/} {\bf 63}(1) 60--81.

\bibitem[{Krajewska and Kopfer(2009)}]{Krajewska2009}
Krajewska, M.A., H.~Kopfer. 2009.
\newblock {Transportation planning in freight forwarding companies}.
\newblock {\it European Journal of Operational Research\/} {\bf 197} 741--751.

\bibitem[{Kramer et~al.(2015)Kramer, Subramanian, Vidal, and
  Cabral}]{Kramer2015}
Kramer, R., A.~Subramanian, T.~Vidal, L.A.F. Cabral. 2015.
\newblock {A matheuristic approach for the pollution-routing problem}.
\newblock {\it European Journal of Operational Research\/} {\bf 243}(2)
  523--539.

\bibitem[{Kwon et~al.(1995)Kwon, Golden, and Wasil}]{Kwon1995}
Kwon, O., B.L. Golden, E.A. Wasil. 1995.
\newblock {Estimating the length of the optimal TSP tour: An empirical study
  using regression and neural networks}.
\newblock {\it Computers {\&} Operations Research\/} {\bf 22}(10) 1039--1046.

\bibitem[{Laporte and Dejax(1989)}]{Laporte1989}
Laporte, G., P.J. Dejax. 1989.
\newblock {Dynamic location-routeing problems}.
\newblock {\it Journal of the Operational Research Society\/} {\bf 40}(5)
  471--482.

\bibitem[{Laporte et~al.(1992)Laporte, Louveaux, and Mercure}]{Laporte1992}
Laporte, G., F.V. Louveaux, H.~Mercure. 1992.
\newblock {The vehicle routing problem with stochastic travel times}.
\newblock {\it Transportation Science\/} {\bf 26}(3) 161--170.

\bibitem[{Laporte et~al.(2018)Laporte, Meunier, and {Wolfler
  Calvo}}]{Laporte2018}
Laporte, G., F.~Meunier, R.~{Wolfler Calvo}. 2018.
\newblock {Shared mobility systems: An updated survey}.
\newblock {\it Annals of Operations Research\/} {\bf 271} 105--126.

\bibitem[{Laporte et~al.(2015)Laporte, Nickel, and {Saldanha da
  Gama}}]{Laporte2015a}
Laporte, G., S.~Nickel, F.~{Saldanha da Gama}, eds. 2015.
\newblock {\it {Location Science}\/}.
\newblock Springer, Berlin.

\bibitem[{Lei et~al.(2015)Lei, Laporte, Liu, and Zhang}]{Lei2015}
Lei, H., G.~Laporte, Y.~Liu, T.~Zhang. 2015.
\newblock {Dynamic design of sales territories}.
\newblock {\it Computers {\&} Operations Research\/} {\bf 56} 84--92.

\bibitem[{Lei et~al.(2016)Lei, Wang, and Laporte}]{Lei2016}
Lei, H., R.~Wang, G.~Laporte. 2016.
\newblock {Solving a multi-objective dynamic stochastic districting and routing
  problem with a co-evolutionary algorithm}.
\newblock {\it Computers {\&} Operations Research\/} {\bf 67} 12--24.

\bibitem[{Li et~al.(2010)Li, Tian, and Leung}]{Li2010}
Li, X., P.~Tian, S.C.H. Leung. 2010.
\newblock {Vehicle routing problems with time windows and stochastic travel and
  service times: Models and algorithm}.
\newblock {\it International Journal of Production Economics\/} {\bf 125}(1)
  137--145.

\bibitem[{Lin et~al.(2017)Lin, Chin, Fu, and Tsui}]{Lin2017}
Lin, M., K.-S. Chin, C.~Fu, K.-L. Tsui. 2017.
\newblock {An effective greedy method for the meals-on-wheels service
  districting problem}.
\newblock {\it Computers {\&} Industrial Engineering\/} {\bf 106} 1--19.

\bibitem[{Liu(2013)}]{Liu2013}
Liu, S. 2013.
\newblock {A hybrid population heuristic for the heterogeneous vehicle routing
  problems}.
\newblock {\it Transportation Research Part E: Logistics and Transportation
  Review\/} {\bf 54} 67--78.

\bibitem[{Lopez et~al.(1998)Lopez, Carter, and Gendreau}]{Lopez1998}
Lopez, L., M.W. Carter, M.~Gendreau. 1998.
\newblock {The hot strip mill production scheduling problem: A tabu search
  approach}.
\newblock {\it European Journal of Operational Research\/} {\bf 106}(2-3)
  317--335.

\bibitem[{Louveaux and Salazar-Gonz{\'{a}}lez(2018)}]{Louveaux2018}
Louveaux, F.V., J.-J. Salazar-Gonz{\'{a}}lez. 2018.
\newblock {Exact approach for the vehicle routing problem with stochastic
  demands and preventive returns}.
\newblock {\it Transportation Science\/} {\bf 52}(6) 1463--1478.

\bibitem[{Lum et~al.(2017)Lum, Cerrone, Golden, and Wasil}]{Lum2017}
Lum, O., C.~Cerrone, B.L. Golden, E.A. Wasil. 2017.
\newblock {Partitioning a street network into compact, balanced, and visually
  appealing routes}.
\newblock {\it Networks\/} {\bf 69}(3) 290--303.

\bibitem[{Luo et~al.(2015)Luo, Qin, Che, and Lim}]{Luo2015}
Luo, Z., H.~Qin, C.~Che, A.~Lim. 2015.
\newblock {On service consistency in multi-period vehicle routing}.
\newblock {\it European Journal of Operational Research\/} {\bf 243}(3)
  731--744.

\bibitem[{Ma et~al.(2012)Ma, Cheang, Lim, Zhang, and Zhu}]{Ma2012}
Ma, H., B.~Cheang, A.~Lim, L.~Zhang, Y.~Zhu. 2012.
\newblock {An investigation into the vehicle routing problem with time windows
  and link capacity constraints}.
\newblock {\it Omega\/} {\bf 40}(3) 336--347.

\bibitem[{Maden et~al.(2009)Maden, Eglese, and Black}]{Maden2009}
Maden, W., R.~Eglese, D.~Black. 2009.
\newblock {Vehicle routing and scheduling with time-varying data: A case
  study}.
\newblock {\it Journal of the Operational Research Society\/} {\bf 61}(3)
  515--522.

\bibitem[{Marques et~al.(2019)Marques, Sadykov, Deschamps, and
  Dupas}]{Marques2019}
Marques, G., R.~Sadykov, J.-C. Deschamps, R.~Dupas. 2019.
\newblock {An improved branch-cut-and-price algorithm for the two-echelon
  capacitated vehicle routing problem}.
\newblock Tech. rep., University of Bordeaux.

\bibitem[{Mart{\'{i}} et~al.(2009)Mart{\'{i}}, Velarde, and Duarte}]{Marti2009}
Mart{\'{i}}, R., J.L.G. Velarde, A.~Duarte. 2009.
\newblock {Heuristics for the bi-objective path dissimilarity problem}.
\newblock {\it Computers {\&} Operations Research\/} {\bf 36}(11) 2905--2912.

\bibitem[{Martinelli and Contardo(2015)}]{Martinelli2015}
Martinelli, R., C.~Contardo. 2015.
\newblock {Exact and heuristic algorithms for capacitated vehicle routing
  problems with quadratic costs structure}.
\newblock {\it INFORMS Journal on Computing\/} {\bf 27}(4) 658--676.

\bibitem[{Matl et~al.(2018)Matl, Hartl, and Vidal}]{Matl2018}
Matl, P., R.F. Hartl, T.~Vidal. 2018.
\newblock {Workload equity in vehicle routing problems: A survey and analysis}.
\newblock {\it Transportation Science\/} {\bf 52}(2) 239--260.

\bibitem[{Matl et~al.(2019)Matl, Hartl, and Vidal}]{Matl2018a}
Matl, P., R.F. Hartl, T.~Vidal. 2019.
\newblock {Workload equity in vehicle routing: The impact of alternative
  workload resources}.
\newblock {\it Computers {\&} Operations Research\/} {\bf 110} 116--129.

\bibitem[{Mendoza et~al.(2009)Mendoza, Medaglia, and Velasco}]{Mendoza2009}
Mendoza, J., A.~Medaglia, N.~Velasco. 2009.
\newblock {An evolutionary-based decision support system for vehicle routing:
  The case of a public utility}.
\newblock {\it Decision Support Systems\/} {\bf 46}(3) 730--742.

\bibitem[{Meng et~al.(2005)Meng, Lee, and Cheu}]{Meng2005}
Meng, Q., D.H. Lee, R.L. Cheu. 2005.
\newblock {Multiobjective vehicle routing and scheduling problem with time
  window constraints in hazardous material transportation}.
\newblock {\it Journal of Transportation Engineering\/} {\bf 131}(9) 699--707.

\bibitem[{Merch{\'{a}}n and Winkenbach(2019)}]{Merchan2019}
Merch{\'{a}}n, D., M.~Winkenbach. 2019.
\newblock {A data-driven extension of continuum approximation approaches to
  predict urban route distances}.
\newblock {\it Networks\/} {\bf 73}(4) 418--433.

\bibitem[{Michallet et~al.(2014)Michallet, Prins, Amodeo, Yalaoui, and
  Vitry}]{Michallet2014}
Michallet, J., C.~Prins, L.~Amodeo, F.~Yalaoui, G.~Vitry. 2014.
\newblock {Multi-start iterated local search for the periodic vehicle routing
  problem with time windows and time spread constraints on services}.
\newblock {\it Computers {\&} Operations Research\/} {\bf 41}(1) 196--207.

\bibitem[{Montoya et~al.(2017)Montoya, Gu{\'{e}}ret, Mendoza, and
  Villegas}]{Montoya2017}
Montoya, A., C.~Gu{\'{e}}ret, J.E. Mendoza, J.G. Villegas. 2017.
\newblock {The electric vehicle routing problem with nonlinear charging
  function}.
\newblock {\it Transportation Research Part B: Methodological\/} {\bf 103}
  87--110.

\bibitem[{Montreuil(2011)}]{Montreuil2011}
Montreuil, B. 2011.
\newblock {Toward a physical Internet: Meeting the global logistics
  sustainability grand challenge}.
\newblock {\it Logistics Research\/} {\bf 3}(2-3) 71--87.

\bibitem[{Morillo and Campos(2014)}]{Morillo2014}
Morillo, C., J.M. Campos. 2014.
\newblock {On-street illegal parking costs in urban areas}.
\newblock {\it Procedia -- Social and Behavioral Sciences\/} {\bf 160}
  342--351.

\bibitem[{Mourad et~al.(2019)Mourad, Puchinger, and Chu}]{Mourad2019}
Mourad, A., J.~Puchinger, C.~Chu. 2019.
\newblock {A survey of models and algorithms for optimizing shared mobility}.
\newblock {\it Transportation Research Part B: Methodological\/} {\bf 123}
  323--346.

\bibitem[{Mufalli et~al.(2012)Mufalli, Batta, and Nagi}]{Mufalli2012}
Mufalli, F., R.~Batta, R.~Nagi. 2012.
\newblock {Simultaneous sensor selection and routing of unmanned aerial
  vehicles for complex mission plans}.
\newblock {\it Computers {\&} Operations Research\/} {\bf 39}(11) 2787--2799.

\bibitem[{M{\"{u}}ller-Hannemann and Weihe(2006)}]{Muller-Hannemann2006}
M{\"{u}}ller-Hannemann, M., K.~Weihe. 2006.
\newblock {On the cardinality of the Pareto set in bicriteria shortest path
  problems}.
\newblock {\it Annals of Operations Research\/} {\bf 147}(1) 269--286.

\bibitem[{Murray and Chu(2015)}]{Murray2015}
Murray, C.C., A.G. Chu. 2015.
\newblock {The flying sidekick traveling salesman problem: Optimization of
  drone-assisted parcel delivery}.
\newblock {\it Transportation Research Part C: Emerging Technologies\/} {\bf
  54} 86--109.

\bibitem[{Neves-Moreira et~al.(2019)Neves-Moreira, Almada-Lobo, Cordeau,
  Guimar{\~{a}}es, and Jans}]{Neves-Moreira2019}
Neves-Moreira, F., B.~Almada-Lobo, J.-F. Cordeau, L.~Guimar{\~{a}}es, R.~Jans.
  2019.
\newblock {Solving a large multi-product production-routing problem with
  delivery time windows}.
\newblock {\it Omega\/} {\bf 86} 154--172.

\bibitem[{Ngueveu et~al.(2010{\natexlab{a}})Ngueveu, Prins, and {Wolfler
  Calvo}}]{Ngueveu2010a}
Ngueveu, S.U., C.~Prins, R.~{Wolfler Calvo}. 2010{\natexlab{a}}.
\newblock {A hybrid tabu search for the $m$-peripatetic vehicle routing problem}.
\newblock V.~Maniezzo, T.~St{\"{u}}tzle, S.~Vo{\ss}, eds., {\it
  Matheuristics\/}. Springer, Boston, MA, 253--266.

\bibitem[{Ngueveu et~al.(2010{\natexlab{b}})Ngueveu, Prins, and {Wolfler
  Calvo}}]{Ngueveu2010c}
Ngueveu, S.U., C.~Prins, R.~{Wolfler Calvo}. 2010{\natexlab{b}}.
\newblock {Lower and upper bounds for the $m$-peripatetic vehicle routing
  problem}.
\newblock {\it 4OR\/} {\bf 8}(4) 387--406.

\bibitem[{Nielsen et~al.(1998)Nielsen, Frederiksen, and Simonsen}]{Nielsen1998}
Nielsen, O.A., R.D. Frederiksen, N.~Simonsen. 1998.
\newblock {Using expert system rules to establish data for intersections and
  turns in road networks}.
\newblock {\it International Transactions in Operational Research\/} {\bf 5}(6)
  569--581.

\bibitem[{Nocerino et~al.(2016)Nocerino, Colorni, Lia, and
  Lu{\`{e}}}]{Nocerino2016}
Nocerino, R., A.~Colorni, F.~Lia, A.~Lu{\`{e}}. 2016.
\newblock {E-bikes and E-scooters for smart logistics: Environmental and
  economic sustainability in pro-E-bike Italian pilots}.
\newblock {\it Transportation Research Procedia\/} {\bf 14} 2362--2371.

\bibitem[{Norstad et~al.(2011)Norstad, Fagerholt, and Laporte}]{Norstad2011}
Norstad, I., K.~Fagerholt, G.~Laporte. 2011.
\newblock {Tramp ship routing and scheduling with speed optimization}.
\newblock {\it Transportation Research Part C: Emerging Technologies\/} {\bf
  19}(5) 853--865.

\bibitem[{Nourinejad and Roorda(2017)}]{Nourinejad2017}
Nourinejad, M., M.J. Roorda. 2017.
\newblock {A continuous approximation model for the fleet composition problem
  on the rectangular grid}.
\newblock {\it OR Spectrum\/} {\bf 39}(2) 373--401.

\bibitem[{Novaes et~al.(2000)Novaes, de~Cursi, and Graciolli}]{Novaes2000}
Novaes, A.G.N., J.E.S. de~Cursi, O.D. Graciolli. 2000.
\newblock {A continuous approach to the design of physical distribution
  systems}.
\newblock {\it Computers {\&} Operations Research\/} {\bf 27}(9) 877--893.

\bibitem[{Nowak et~al.(2008)Nowak, Ergun, and {White III}}]{Nowak2008}
Nowak, M., {\"{O}}.~Ergun, C.C. {White III}. 2008.
\newblock {Pickup and delivery with split loads}.
\newblock {\it Transportation Science\/} {\bf 42}(1) 32--43.

\bibitem[{Orda and Rom(1990)}]{Orda1990}
Orda, A., R~Rom. 1990.
\newblock {Shortest-path and minimum-delay algorithms in networks with
  time-dependent edge-length}.
\newblock {\it Journal of the Association for Computing Machinery\/} {\bf
  37}(3) 607--625.

\bibitem[{Orlis et~al.(2019)Orlis, Lagan{\'{a}}, Dullaert, and
  Vigo}]{Orlis2019}
Orlis, C., D.~Lagan{\'{a}}, W.~Dullaert, D.~Vigo. 2019.
\newblock {Distribution with quality of service considerations: The capacitated
  routing problem with profits and service level requirements}.
\newblock {\it Omega, Articles in Advance\/} .

\bibitem[{Ouyang and Daganzo(2006)}]{Ouyang2006}
Ouyang, Y., C.F. Daganzo. 2006.
\newblock {Discretization and validation of the continuum approximation scheme
  for terminal system design}.
\newblock {\it Transportation Science\/} {\bf 40}(1) 89--98.

\bibitem[{{Padilla Tinoco} et~al.(2017){Padilla Tinoco}, Creemers, and
  Boute}]{Tinoco2017}
{Padilla Tinoco}, S.V., S.~Creemers, R.N. Boute. 2017.
\newblock {Collaborative shipping under different cost-sharing agreements}.
\newblock {\it European Journal of Operational Research\/} {\bf 263}(3)
  827--837.

\bibitem[{Pantuso et~al.(2014)Pantuso, Fagerholt, and Hvattum}]{Pantuso2014}
Pantuso, G., K.~Fagerholt, L.M. Hvattum. 2014.
\newblock {A survey on maritime fleet size and mix problems}.
\newblock {\it European Journal of Operational Research\/} {\bf 235}(2)
  341--349.

\bibitem[{Papageorgiou et~al.(2014)Papageorgiou, Nemhauser, Sokol, Cheon, and
  Keha}]{Papageorgiou2014}
Papageorgiou, D.J., G.L. Nemhauser, J.~Sokol, M.-S. Cheon, A.B. Keha. 2014.
\newblock {MIRPLib -- A library of maritime inventory routing problem
  instances: Survey, core model, and benchmark results}.
\newblock {\it European Journal of Operational Research\/} {\bf 235}(2)
  350--366.

\bibitem[{Paquette et~al.(2012)Paquette, Bellavance, Cordeau, and
  Laporte}]{Paquette2012}
Paquette, J., F.~Bellavance, J.-F. Cordeau, G.~Laporte. 2012.
\newblock {Measuring quality of service in dial-a-ride operations: The case of
  a Canadian city}.
\newblock {\it Transportation\/} {\bf 39}(3) 539--564.

\bibitem[{Paraskevopoulos et~al.(2017)Paraskevopoulos, Laporte, Repoussis, and
  Tarantilis}]{Paraskevopoulos2017}
Paraskevopoulos, D.C., G.~Laporte, P.P. Repoussis, C.D. Tarantilis. 2017.
\newblock {Resource constrained routing and scheduling: Review and research
  prospects}.
\newblock {\it European Journal of Operational Research\/} {\bf 263}(3)
  737--754.

\bibitem[{Park and Kim(2010)}]{Park2010}
Park, J., B.-I. Kim. 2010.
\newblock {The school bus routing problem: A review}.
\newblock {\it European Journal of Operational Research\/} {\bf 202}(2)
  311--319.

\bibitem[{Parragh and Doerner(2018)}]{Parragh2018}
Parragh, S.N., K.F. Doerner. 2018.
\newblock {Solving routing problems with pairwise synchronization constraints}.
\newblock {\it Central European Journal of Operations Research\/} {\bf 26}(2)
  443--464.

\bibitem[{Pasha et~al.(2016)Pasha, Hoff, and Hvattum}]{Pasha2016}
Pasha, U., A.~Hoff, L.M. Hvattum. 2016.
\newblock {Simple heuristics for the multi-period fleet size and mix vehicle
  routing problem}.
\newblock {\it INFOR: Information Systems and Operational Research\/} {\bf
  54}(2) 97--120.

\bibitem[{Paterson et~al.(2011)Paterson, Kiesm{\"{u}}ller, Teunter, and
  Glazebrook}]{Paterson2011}
Paterson, C., G.~Kiesm{\"{u}}ller, R.~Teunter, K.~Glazebrook. 2011.
\newblock {Inventory models with lateral transshipments: A review}.
\newblock {\it European Journal of Operational Research\/} {\bf 210}(2)
  125--136.

\bibitem[{Pelletier et~al.(2016)Pelletier, Jabali, and Laporte}]{Pelletier2016}
Pelletier, S., O.~Jabali, G.~Laporte. 2016.
\newblock {Goods distribution with electric vehicles: Review and research
  perspectives}.
\newblock {\it Transportation Science\/} {\bf 50}(1) 3--22.

\bibitem[{Pelletier et~al.(2018)Pelletier, Jabali, and Laporte}]{Pelletier2018}
Pelletier, S., O.~Jabali, G.~Laporte. 2018.
\newblock {Charge scheduling for electric freight vehicles}.
\newblock {\it Transportation Research Part B: Methodological\/} {\bf 115}
  246--269.

\bibitem[{Penna et~al.(2019)Penna, Subramanian, Ochi, Vidal, and
  Prins}]{Penna2019}
Penna, P.H.V., A.~Subramanian, L.S. Ochi, T.~Vidal, C.~Prins. 2019.
\newblock {A hybrid heuristic for a broad class of vehicle routing problems
  with heterogeneous fleet}.
\newblock {\it Annals of Operations Research\/} {\bf 273}(1-2) 5--74.

\bibitem[{Perrier et~al.(2008)Perrier, Langevin, and Amaya}]{Perrier2008}
Perrier, N., A.~Langevin, C.-A. Amaya. 2008.
\newblock {Vehicle routing for urban snow plowing operations}.
\newblock {\it Transportation Science\/} {\bf 42}(1) 44--56.

\bibitem[{Pessoa et~al.(2018{\natexlab{a}})Pessoa, Sadykov, and
  Uchoa}]{Pessoa2018}
Pessoa, A., R.~Sadykov, E.~Uchoa. 2018{\natexlab{a}}.
\newblock {Enhanced branch-cut-and-price algorithm for heterogeneous fleet
  vehicle routing problems}.
\newblock {\it European Journal of Operational Research\/} {\bf 270}(2)
  530--543.

\bibitem[{Pessoa et~al.(2019)Pessoa, Sadykov, Uchoa, and
  Vanderbeck}]{Pessoa2019}
Pessoa, A., R.~Sadykov, E.~Uchoa, F.~Vanderbeck. 2019.
\newblock {A generic exact solver for vehicle routing and related problems}.
\newblock {\it IPCO\/}. Ann Arbor, Michigan, USA.

\bibitem[{Pessoa et~al.(2018{\natexlab{b}})Pessoa, Poss, Sadykov, and
  Vanderbeck}]{Pessoa2018b}
Pessoa, A.A., M.~Poss, R.~Sadykov, F.~Vanderbeck. 2018{\natexlab{b}}.
\newblock {Branch-and-cut-and-price for the robust capacitated vehicle routing
  problem with knapsack uncertainty}.
\newblock Tech. rep., Cadernos do LOGIS-UFF 2018-01, Niteroi.

\bibitem[{Pillac et~al.(2013)Pillac, Gendreau, Gu{\'{e}}ret, and
  Medaglia}]{Pillac2013}
Pillac, V., M.~Gendreau, C.~Gu{\'{e}}ret, A.L. Medaglia. 2013.
\newblock {A review of dynamic vehicle routing problems}.
\newblock {\it European Journal of Operational Research\/} {\bf 225}(1) 1--11.

\bibitem[{Poikonen et~al.(2017)Poikonen, Wang, and Golden}]{Poikonen2017}
Poikonen, S., X.~Wang, B.L. Golden. 2017.
\newblock {The vehicle routing problem with drones: Extended models and
  connections}.
\newblock {\it Networks\/} {\bf 70}(1) 34--43.

\bibitem[{Pollaris et~al.(2015)Pollaris, Braekers, Caris, Janssens, and
  Limbourg}]{Pollaris2015}
Pollaris, H., K.~Braekers, A.~Caris, G.K. Janssens, S.~Limbourg. 2015.
\newblock {Vehicle routing problems with loading constraints: State-of-the-art
  and future directions}.
\newblock {\it OR Spectrum\/} {\bf 37}(2) 297--330.

\bibitem[{Pollaris et~al.(2017)Pollaris, Braekers, Caris, Janssens, and
  Limbourg}]{Pollaris2017}
Pollaris, H., K.~Braekers, A.~Caris, G.K. Janssens, S.~Limbourg. 2017.
\newblock {Iterated local search for the capacitated vehicle routing problem
  with sequence-based pallet loading and axle weight constraints}.
\newblock {\it Networks\/} {\bf 69}(3) 304--316.

\bibitem[{Poot et~al.(2002)Poot, Kant, and Wagelmans}]{Poot2002}
Poot, A., G.~Kant, A.~P.M. Wagelmans. 2002.
\newblock {A savings based method for real-life vehicle routing problems}.
\newblock {\it Journal of the Operational Research Society\/} {\bf 53}(1)
  57--68.

\bibitem[{Prescott-Gagnon et~al.(2010)Prescott-Gagnon, Desaulniers, Drexl, and
  Rousseau}]{Prescott-Gagnon2010}
Prescott-Gagnon, E., G.~Desaulniers, M.~Drexl, L.-M. Rousseau. 2010.
\newblock {European driver rules in vehicle routing with time windows}.
\newblock {\it Transportation Science\/} {\bf 44}(4) 455--473.

\bibitem[{Prodhon and Prins(2014)}]{Prodhon2014}
Prodhon, C., C.~Prins. 2014.
\newblock {A survey of recent research on location-routing problems}.
\newblock {\it European Journal of Operational Research\/} {\bf 238} 1--17.

\bibitem[{Qiu et~al.(2019)Qiu, Qiao, and Pardalos}]{Qiu2019}
Qiu, Y., J.~Qiao, P.M. Pardalos. 2019.
\newblock {Optimal production, replenishment, delivery, routing and inventory
  management policies for products with perishable inventory}.
\newblock {\it Omega\/} {\bf 82} 193--204.

\bibitem[{Rasmussen et~al.(2012)Rasmussen, Justesen, Dohn, and
  Larsen}]{Rasmussen2012}
Rasmussen, M.S., T.~Justesen, A.~Dohn, J.~Larsen. 2012.
\newblock {The home care crew scheduling problem: Preference-based visit
  clustering and temporal dependencies}.
\newblock {\it European Journal of Operational Research\/} {\bf 219}(3)
  598--610.

\bibitem[{Reinhardt et~al.(2016)Reinhardt, Jepsen, and
  Pisinger}]{Reinhardt2016}
Reinhardt, L.B., M.K. Jepsen, D.~Pisinger. 2016.
\newblock {The edge set cost of the vehicle routing problem with time windows}.
\newblock {\it Transportation Science\/} {\bf 50}(2) 694--707.

\bibitem[{Rincon-Garcia et~al.(2018)Rincon-Garcia, Waterson, and
  Cherrett}]{Rincon2018}
Rincon-Garcia, N., B.J. Waterson, T.J. Cherrett. 2018.
\newblock {Requirements from vehicle routing software: Perspectives from
  literature, developers and the freight industry}.
\newblock {\it Transport Reviews\/} {\bf 38}(1) 117--138.

\bibitem[{R{\'{i}}os-Mercado and Fern{\'{a}}ndez(2009)}]{Rios2009}
R{\'{i}}os-Mercado, R.S., E.~Fern{\'{a}}ndez. 2009.
\newblock {A reactive GRASP for a commercial territory design problem with
  multiple balancing requirements}.
\newblock {\it Computers {\&} Operations Research\/} {\bf 36}(3) 755--776.

\bibitem[{Rosenfield et~al.(1992)Rosenfield, Engelstein, and
  Feigenbaum}]{Rosenfield1992}
Rosenfield, D.B., I.~Engelstein, D.~Feigenbaum. 1992.
\newblock {An application of sizing service territories}.
\newblock {\it European Journal of Operational Research\/} {\bf 63}(2)
  164--172.

\bibitem[{Rossit et~al.(2019)Rossit, Vigo, Tohm{\'{e}}, and
  Frutos}]{Rossit2019}
Rossit, D.G., D.~Vigo, F.~Tohm{\'{e}}, M.~Frutos. 2019.
\newblock {Visual attractiveness in routing problems: A review}.
\newblock {\it Computers {\&} Operations Research\/} {\bf 103} 13--34.

\bibitem[{Rostami et~al.(2017)Rostami, Desaulniers, Errico, and
  Lodi}]{Rostami2017}
Rostami, B., G.~Desaulniers, F.~Errico, A.~Lodi. 2017.
\newblock {The vehicle routing problem with stochastic and correlated travel
  times}.
\newblock Tech. rep., Polytechnique Montr{\'{e}}al.

\bibitem[{Sahin et~al.(2013)Sahin, {\c{C}}avuşlar, {\"{O}}ncan, Sahin, and
  Aksu}]{Sahin2013}
Sahin, M., G.~{\c{C}}avuşlar, T.~{\"{O}}ncan, G.~Sahin, D.T. Aksu. 2013.
\newblock {An efficient heuristic for the multi-vehicle one-to-one pickup and
  delivery problem with split loads}.
\newblock {\it Transportation Research Part C: Emerging Technologies\/} {\bf
  27} 169--188.

\bibitem[{Salavati-Khoshghalb et~al.(2019)Salavati-Khoshghalb, Gendreau,
  Jabali, and Rei}]{Salavati-Khoshghalb2019}
Salavati-Khoshghalb, M., M.~Gendreau, O.~Jabali, W.~Rei. 2019.
\newblock {An exact algorithm to solve the vehicle routing problem with
  stochastic demands under an optimal restocking policy}.
\newblock {\it European Journal of Operational Research\/} {\bf 273}(1)
  175--189.

\bibitem[{Schiffer et~al.(2017)Schiffer, Laporte, Schneider, and
  Walther}]{Schiffer2017c}
Schiffer, M., G.~Laporte, M.~Schneider, G.~Walther. 2017.
\newblock {The impact of synchronizing driver breaks and recharging operations
  for electric vehicles}.
\newblock Tech. rep., RWTH, Aachen.

\bibitem[{Schiffer et~al.(2019)Schiffer, Schneider, Walther, and
  Laporte}]{Schiffer2019}
Schiffer, M., M.~Schneider, G.~Walther, G.~Laporte. 2019.
\newblock {Vehicle routing and location-routing with intermediate stops: A
  review}.
\newblock {\it Transportation Science\/} {\bf 53}(2) 319--343.

\bibitem[{Schneider and Drexl(2017)}]{Schneider2017c}
Schneider, M., M.~Drexl. 2017.
\newblock {A survey of the standard location-routing problem}.
\newblock {\it Annals of Operations Research\/} {\bf 259}(1-2) 389--414.

\bibitem[{Schneider and L{\"{o}}ffler(2019)}]{Schneider2019}
Schneider, M., M.~L{\"{o}}ffler. 2019.
\newblock {Large composite neighborhoods for the capacitated location-routing
  problem}.
\newblock {\it Transportation Science\/} {\bf 53}(1) 301--318.

\bibitem[{Schneider et~al.(2015)Schneider, Stenger, Schwahn, and
  Vigo}]{Schneider2015}
Schneider, M., A.~Stenger, F.~Schwahn, D.~Vigo. 2015.
\newblock {Territory-based vehicle routing in the presence of time-window
  constraints}.
\newblock {\it Transportation Science\/} {\bf 49}(4) 732--751.

\bibitem[{Shah et~al.(2011)Shah, Bao, Lu, and Chen}]{Shah2011}
Shah, S., F.~Bao, C.-T. Lu, I.-R. Chen. 2011.
\newblock {Crowdsafe: Crowd sourcing of crime incidents and safe routing on
  mobile devices}.
\newblock {\it 9th ACM SIGSPATIAL International Conference on Advances in
  Geographic Information Systems\/}. 521--524.

\bibitem[{Shao et~al.(2014)Shao, Kulik, Tanin, and Guo}]{Shao2014}
Shao, J., L.~Kulik, E.~Tanin, L.~Guo. 2014.
\newblock {Travel distance versus navigation complexity: A study on different
  spatial queries on road networks}.
\newblock {\it 23rd ACM International Conference on Conference on Information
  and Knowledge Management\/}. 1791--1794.

\bibitem[{Shelbourne et~al.(2017)Shelbourne, Battarra, and
  Potts}]{Shelbourne2017}
Shelbourne, C., M.~Battarra, C.N. Potts. 2017.
\newblock {The vehicle routing problem with release and due dates}.
\newblock {\it INFORMS Journal on Computing\/} {\bf 29}(4) 705--723.

\bibitem[{Silva et~al.(2012)Silva, Subramanian, Vidal, and Ochi}]{Silva2012}
Silva, M.M., A.~Subramanian, T.~Vidal, L.S. Ochi. 2012.
\newblock {A simple and effective metaheuristic for the minimum latency
  problem}.
\newblock {\it European Journal of Operational Research\/} {\bf 221}(3)
  513--520.

\bibitem[{Skiera and Albers(1998)}]{Skiera1998}
Skiera, B., S.~Albers. 1998.
\newblock {COSTA: Contribution optimizing sales territory alignment}.
\newblock {\it Marketing Science\/} {\bf 17}(3) 196--213.

\bibitem[{Smilowitz et~al.(2013)Smilowitz, Nowak, and Jiang}]{Smilowitz2013}
Smilowitz, K., M.~Nowak, T.~Jiang. 2013.
\newblock {Workforce management in periodic delivery operations}.
\newblock {\it Transportation Science\/} {\bf 47}(2) 214--230.

\bibitem[{Solomon(1987)}]{Solomon1987}
Solomon, M.M. 1987.
\newblock {Algorithms for the vehicle routing and scheduling problems with time
  window constraints}.
\newblock {\it Operations Research\/} {\bf 35}(2) 254--265.

\bibitem[{Song and Savelsbergh(2007)}]{Song2007}
Song, J.-H., M.W.P. Savelsbergh. 2007.
\newblock {Performance measurement for inventory routing}.
\newblock {\it Transportation Science\/} {\bf 41}(1) 44--54.

\bibitem[{Soriano et~al.(2019)Soriano, Vidal, Gansterer, and
  Doerner}]{Soriano2019}
Soriano, A., T.~Vidal, M.~Gansterer, K.~Doerner. 2019.
\newblock {The vehicle routing problem with arrival time diversification on a
  multigraph}.
\newblock Tech. rep., University of Vienna, Austria.

\bibitem[{Soysal et~al.(2015)Soysal, Bloemhof-Ruwaard, and
  Bektaş}]{Soysal2015}
Soysal, M., J.M. Bloemhof-Ruwaard, T.~Bektaş. 2015.
\newblock {The time-dependent two-echelon capacitated vehicle routing problem
  with environmental considerations}.
\newblock {\it International Journal of Production Economics\/} {\bf 164}
  366--378.

\bibitem[{Spliet and Dekker(2016)}]{Spliet2016}
Spliet, R., R.~Dekker. 2016.
\newblock {The driver assignment vehicle routing problem}.
\newblock {\it Networks\/} {\bf 68}(3) 212--223.

\bibitem[{Spliet and Gabor(2015)}]{Spliet2015b}
Spliet, R., A.F. Gabor. 2015.
\newblock {The time window assignment vehicle routing problem}.
\newblock {\it Transportation Science\/} {\bf 49}(4) 721--731.

\bibitem[{St{\aa}lhane et~al.(2012)St{\aa}lhane, Rakke, Moe, Andersson,
  Christiansen, and Fagerholt}]{Stalhane2012}
St{\aa}lhane, M., J.G. Rakke, C.R. Moe, H.~Andersson, M.~Christiansen,
  K.~Fagerholt. 2012.
\newblock {A construction and improvement heuristic for a liquefied natural gas
  inventory routing problem}.
\newblock {\it Computers {\&} Industrial Engineering\/} {\bf 62}(1) 245--255.

\bibitem[{Stenger et~al.(2013{\natexlab{a}})Stenger, Schneider, and
  Goeke}]{Stenger2013}
Stenger, A., M.~Schneider, D.~Goeke. 2013{\natexlab{a}}.
\newblock {The prize-collecting vehicle routing problem with single and
  multiple depots and non-linear cost}.
\newblock {\it EURO Journal on Transportation and Logistics\/} {\bf 2}(1-2)
  57--87.

\bibitem[{Stenger et~al.(2013{\natexlab{b}})Stenger, Vigo, Enz, and
  Schwind}]{Stenger2013a}
Stenger, A., D.~Vigo, S.~Enz, M.~Schwind. 2013{\natexlab{b}}.
\newblock {An adaptive variable neighborhood search algorithm for a vehicle
  routing problem arising in small package shipping}.
\newblock {\it Transportation Science\/} {\bf 47}(1) 64--80.

\bibitem[{Sungur et~al.(2008)Sungur, Ord{\'{o}}{\~{n}}ez, and
  Dessouky}]{Sungur2008}
Sungur, I., F.~Ord{\'{o}}{\~{n}}ez, M.~Dessouky. 2008.
\newblock {A robust optimization approach for the capacitated vehicle routing
  problem with demand uncertainty}.
\newblock {\it IIE Transactions\/} {\bf 40}(5) 509--523.

\bibitem[{Sungur et~al.(2010)Sungur, Ren, Ord{\'{o}}{\~{n}}ez, Dessouky, and
  Zhong}]{Sungur2010}
Sungur, I., Y.~Ren, F.~Ord{\'{o}}{\~{n}}ez, M.~Dessouky, H.~Zhong. 2010.
\newblock {A model and algorithm for the courier delivery problem with
  uncertainty}.
\newblock {\it Transportation Science\/} {\bf 44}(2) 193--205.

\bibitem[{Talarico et~al.(2015)Talarico, S{\"{o}}rensen, and
  Springael}]{Talarico2015}
Talarico, L., K.~S{\"{o}}rensen, J.~Springael. 2015.
\newblock {Metaheuristics for the risk-constrained cash-in-transit vehicle
  routing problem}.
\newblock {\it European Journal of Operational Research\/} {\bf 244}(2)
  457--470.

\bibitem[{Tang and Miller-Hooks(2006)}]{Tang2006}
Tang, H., E.~Miller-Hooks. 2006.
\newblock {Interactive heuristic for practical vehicle routing problem with
  solution shape constraints}.
\newblock {\it Transportation Research Record\/} {\bf 1964}(1) 9--18.

\bibitem[{Tang et~al.(2010)Tang, Kong, Lau, and Ip}]{Tang2010}
Tang, J., Y.~Kong, H.~Lau, A.W.H. Ip. 2010.
\newblock {A note on ``Efficient feasibility testing for dial-a-ride
  problems''}.
\newblock {\it Operations Research Letters\/} {\bf 38}(5) 405--407.

\bibitem[{Tarantilis and Kiranoudis(2001)}]{Tarantilis2001}
Tarantilis, C.D., C.T. Kiranoudis. 2001.
\newblock {Using the vehicle routing problem for the transportation of
  hazardous materials}.
\newblock {\it Operational Research. An International Journal\/} {\bf 1}(1)
  67--78.

\bibitem[{Tarantilis et~al.(2012)Tarantilis, Stavropoulou, and
  Repoussis}]{Tarantilis2012}
Tarantilis, C.D., F.~Stavropoulou, P.P. Repoussis. 2012.
\newblock {A template-based tabu search algorithm for the consistent vehicle
  routing problem}.
\newblock {\it Expert Systems with Applications\/} {\bf 39}(4) 4233--4239.

\bibitem[{Taslimi et~al.(2017)Taslimi, Batta, and Kwon}]{Taslimi2017}
Taslimi, M., R.~Batta, C.~Kwon. 2017.
\newblock {A comprehensive modeling framework for hazmat network design, hazmat
  response team location, and equity of risk}.
\newblock {\it Computers {\&} Operations Research\/} {\bf 79} 119--130.

\bibitem[{Toffolo et~al.(2019)Toffolo, Vidal, and Wauters}]{Toffolo2019}
Toffolo, T.A.M., T.~Vidal, T.~Wauters. 2019.
\newblock {Heuristics for vehicle routing problems: Sequence or set
  optimization?}
\newblock {\it Computers {\&} Operations Research\/} {\bf 105} 118--131.

\bibitem[{Toth and Vigo(2014)}]{Toth2014}
Toth, P., D.~Vigo, eds. 2014.
\newblock {\it {Vehicle Routing: Problems, Methods, and Applications}\/}.
\newblock 2nd ed. Society for Industrial and Applied Mathematics, Philadelphia.

\bibitem[{Toumazis and Kwon(2013)}]{Toumazis2013}
Toumazis, I., C.~Kwon. 2013.
\newblock {Routing hazardous materials on time-dependent networks using
  conditional value-at-risk}.
\newblock {\it Transportation Research Part C: Emerging Technologies\/} {\bf
  37} 73--92.

\bibitem[{Uchoa et~al.(2017)Uchoa, Pecin, Pessoa, Poggi, Subramanian, and
  Vidal}]{Uchoa2017}
Uchoa, E., D.~Pecin, A.~Pessoa, M.~Poggi, A.~Subramanian, T.~Vidal. 2017.
\newblock {New benchmark instances for the capacitated vehicle routing
  problem}.
\newblock {\it European Journal of Operational Research\/} {\bf 257}(3)
  845--858.

\bibitem[{van Anholt et~al.(2016)van Anholt, Coelho, Laporte, and
  Vis}]{VanAnholt2016}
van Anholt, R.G., L.C. Coelho, G.~Laporte, I.F.A. Vis. 2016.
\newblock {An inventory-routing problem with pickups and deliveries arising in
  the replenishment of automated teller machines}.
\newblock {\it Transportation Science\/} {\bf 50}(3) 1077--1091.

\bibitem[{Vanhove and Fack(2012)}]{Vanhove2012}
Vanhove, S., V.~Fack. 2012.
\newblock {Route planning with turn restrictions: A computational experiment}.
\newblock {\it Operations Research Letters\/} {\bf 40}(5) 342--348.

\bibitem[{Vansteenwegen and Souffriau(2009)}]{Vansteenwegen2009c}
Vansteenwegen, P, W~Souffriau. 2009.
\newblock {Metaheuristics for tourist trip planning}.
\newblock K.~S{\"{o}}rensen, M.~Sevaux, W.~Habenicht, M.J. Geiger, eds., {\it
  Metaheuristics in the Service Industry\/}. LNEMS, Springer Berlin Heidelberg,
  15--31.

\bibitem[{Vidal(2017)}]{Vidal2017b}
Vidal, T. 2017.
\newblock {Node, edge, arc routing and turn penalties: Multiple problems -- One
  neighborhood extension}.
\newblock {\it Operations Research\/} {\bf 65}(4) 992--1010.

\bibitem[{Vidal et~al.(2013)Vidal, Crainic, Gendreau, and Prins}]{Vidal2012a}
Vidal, T., T.G. Crainic, M.~Gendreau, C.~Prins. 2013.
\newblock {Heuristics for multi-attribute vehicle routing problems: A survey
  and synthesis}.
\newblock {\it European Journal of Operational Research\/} {\bf 231}(1) 1--21.

\bibitem[{Vidal et~al.(2014)Vidal, Crainic, Gendreau, and Prins}]{Vidal2012b}
Vidal, T., T.G. Crainic, M.~Gendreau, C.~Prins. 2014.
\newblock {A unified solution framework for multi-attribute vehicle routing
  problems}.
\newblock {\it European Journal of Operational Research\/} {\bf 234}(3)
  658--673.

\bibitem[{Vidal et~al.(2015)Vidal, Crainic, Gendreau, and Prins}]{Vidal2015b}
Vidal, T., T.G. Crainic, M.~Gendreau, C.~Prins. 2015.
\newblock {Timing problems and algorithms: Time decisions for sequences of
  activities}.
\newblock {\it Networks\/} {\bf 65}(2) 102--128.

\bibitem[{Vidal et~al.(2019)Vidal, Martinelli, Pham, and H{\`{a}}}]{Vidal2019b}
Vidal, T., R.~Martinelli, T.A. Pham, M.H. H{\`{a}}. 2019.
\newblock {The capacitated arc routing problem with time-dependent travel times
  and paths}.
\newblock Tech. rep., PUC-Rio.

\bibitem[{Villegas et~al.(2013)Villegas, Prins, Prodhon, Medaglia, and
  Velasco}]{Villegas2013}
Villegas, J.G., C.~Prins, C.~Prodhon, A.L. Medaglia, N.~Velasco. 2013.
\newblock {A matheuristic for the truck and trailer routing problem}.
\newblock {\it European Journal of Operational Research\/} {\bf 230}(2)
  231--244.

\bibitem[{Wang et~al.(2018)Wang, Fagerholt, and Wallace}]{Wang2018}
Wang, X., K.~Fagerholt, S.W. Wallace. 2018.
\newblock {Planning for charters: A stochastic maritime fleet composition and
  deployment problem}.
\newblock {\it Omega\/} {\bf 79} 54--66.

\bibitem[{Wong and Beasley(1984)}]{Wong1984a}
Wong, J.F., J.E. Beasley. 1984.
\newblock {Vehicle routing using fixed delivery areas}.
\newblock {\it Omega\/} {\bf 12}(6) 591--600.

\bibitem[{Xie et~al.(2017)Xie, Potts, and Bektaş}]{Xie2017}
Xie, F., C.N. Potts, T.~Bektaş. 2017.
\newblock {Iterated local search for workforce scheduling and routing
  problems}.
\newblock {\it Journal of Heuristics\/} {\bf 23}(6) 471--500.

\bibitem[{Xie and Ouyang(2015)}]{Xie2015}
Xie, W., Y.~Ouyang. 2015.
\newblock {Optimal layout of transshipment facility locations on an infinite
  homogeneous plane}.
\newblock {\it Transportation Research Part B: Methodological\/} {\bf 75}
  74--88.

\bibitem[{Yang et~al.(2000)Yang, Mathur, and Ballou}]{Yang2000b}
Yang, W.-H., K.~Mathur, R.H. Ballou. 2000.
\newblock {Stochastic vehicle routing problem with restocking}.
\newblock {\it Transportation Science\/} {\bf 34}(1) 99--112.

\bibitem[{Zajac(2016)}]{Zajac2016}
Zajac, S. 2016.
\newblock {The bi-objective $k$-dissimilar vehicle routing problem}.
\newblock A.~Paias, M.~Ruthmair, S.~Vo{\ss}, eds., {\it Computational
  Logistics\/}. Springer, Berlin Heidelberg, 306--320.

\bibitem[{Zhang et~al.(2019)Zhang, Baldacci, Sim, and Tang}]{Zhang2019a}
Zhang, Y., R.~Baldacci, M.~Sim, J.~Tang. 2019.
\newblock {Routing optimization with time windows under uncertainty}.
\newblock {\it Mathematical Programming\/} {\bf 175}(1-2) 263--305.

\bibitem[{Zhong et~al.(2007)Zhong, Hall, and Dessouky}]{Zhong2007}
Zhong, H., R.~W. Hall, M.~Dessouky. 2007.
\newblock {Territory planning and vehicle dispatching with driver learning}.
\newblock {\it Transportation Science\/} {\bf 41}(1) 74--89.

\bibitem[{Zografos(2004)}]{Zografos2004}
Zografos, K. 2004.
\newblock {A heuristic algorithm for solving hazardous materials distribution
  problems}.
\newblock {\it European Journal of Operational Research\/} {\bf 152}(2)
  507--519.

\end{thebibliography}

\end{document}